\documentclass[reprint,amsmath,amssymb,aps]{revtex4-2}

\usepackage{graphicx}
\usepackage{dcolumn}
\usepackage{bm}
\usepackage[utf8]{inputenc}
\usepackage[T1]{fontenc}
\usepackage{mathptmx}
\usepackage{etoolbox}
\usepackage{xcolor}
\usepackage{amsmath}
\usepackage{amssymb}
\usepackage{breqn}
\usepackage{mathtools}
\usepackage{stmaryrd}

\makeatletter
\let\cat@comma@active\@empty
\makeatother

\begin{document}

\preprint{APS/123-QED}

\title{Frequency-dependent hydrodynamic finite size correction in molecular simulations\\reveals the long-time hydrodynamic tail}

\author{Laura Scalfi}
\author{Domenico Vitali}
\author{Henrik Kiefer}
\author{Roland R. Netz}%
\email{rnetz@physik.fu-berlin.de}
\affiliation{Fachbereich Physik, Freie Universit\"at Berlin, Arnimallee 14, 14195 Berlin, Germany}

\date{\today}

\begin{abstract}
Finite-size effects are challenging in molecular dynamics simulations because they have significant effects on computed static and dynamic properties, in particular diffusion constants, friction coefficients and time- or frequency-dependent response functions. We investigate the influence of periodic boundary conditions on the velocity autocorrelation function and the frequency-dependent friction of a particle in a fluid and show that the long-time behavior (starting at the picosecond timescale) is significantly affected. We develop an analytical correction allowing to subtract the periodic boundary condition effects. By this we unmask the power-law long-time tails of the memory kernel and the velocity autocorrelation function in liquid water and a Lennard-Jones fluid from rather small simulation boxes.
\end{abstract}

\maketitle

With the progress in computational power, molecular dynamics (MD) simulations have become an essential tool to investigate the properties of matter at the microscopic scale. The accessible length and time scales have not ceased to increase and with them the accuracy of the simulations. 
However, simulation boxes are still limited to the nanometer scale and delimited for example by repulsive walls or more commonly by periodic boundary conditions (PBC).
This finite system size introduces constraints and interactions with the walls or other replicas, and yields a multitude of static finite-size effects in various fields, including surface tension, stress tensors and capillary waves~\cite{gelfand_finite-size_1990,velazquez_finite-size_2006,stukan_finite_2002}, nucleation~\cite{wedekind_finite-size_2006}, phase transitions~\cite{binder_finite_1987,borgs_finite-size_1992} and critical phenomena~\cite{fisher_scaling_1972,ballesteros_finite_1996}. 
Periodicity is particularly relevant for electrostatic interactions~\cite{fraser_finite-size_1996}: for inhomogeneous systems, significant dipole interactions between replicas occur, which are tackled by the Yeh-Berkowitz dipole correction~\cite{figueirido_finitesize_1995,yeh_ewald_1999}. 
Dynamic properties also present finite-size effects due to hydrodynamic interactions, which have mostly been investigated in the stationary limit, for example for the thermal conductivity~\cite{chantrenne_finite_2004,wei_finite-size_2019}, the diffusion coefficient $D$~\cite{dunweg_molecular_1993,klauda_dynamical_2006,dos_santos_self-diffusion_2020} or the friction coefficient $\gamma = k_BT/D$ (with $k_BT$ the thermal energy).
Recently, research has shifted towards time- (and frequency-) dependent response phenomena to characterize transient and non-equilibrium dynamics in complex systems. Finite-size effects have been found in polymer, glass or supercooled fluid dynamics, by investigating the time-dependent dynamic structure factor~\cite{dunweg_microscopic_1991,horbach_finite_1996,kim_apparent_2000}, but studies of the effect of PBC on transient response functions are rare~\cite{asta_transient_2017}. 

In this Letter, we investigate the finite-size dependence of the velocity autocorrelation function (VACF) and of the time-dependent friction function $\Gamma(t)$ (or memory kernel), that quantifies the non-Markovian friction effects in generalized Langevin equations (GLE). For illustrating our general method, we address the simple case of the position fluctuations of a tagged molecule in a fluid. The associated memory kernels have recently been investigated using molecular dynamics simulations to bridge the gap between macroscopic hydrodynamics, where the particle is subject to friction, and Hamiltonian dynamics~\cite{lesnicki_molecular_2016,straube_rapid_2020}. Simulations were compared to hydrodynamic predictions of the friction experienced by a sphere in a fluid.
Indeed, hydrodynamic and mode coupling theories predict a negative long-time friction kernel with an asymptotic power-law decay~\cite{alder_velocity_1967,alder_decay_1970,zwanzig_nonequilibrium_2001,corngold_behavior_1972,lesnicki_molecular_2016}
\begin{equation}
    \Gamma_{\rm tail}(t) = - \frac{2 \gamma^2}{3 \rho} \left[4 \pi \left(D + \frac{\eta}{\rho}\right)t\right]^{-3/2}\,, \label{eq:gamma_tail}
\end{equation}
with $\rho$ the mass density and $\eta$ the shear viscosity of the fluid. 
The contribution proportional to the diffusion coefficient $D$ comes from the particle diffusion and is often negligible with respect to the kinematic viscosity $\eta / \rho$ (see Appendix~\ref{app:sim}). 
Such long-time decay is reflected in the VACF $C^{vv}$, for which a positive $t^{-3/2}$ decay is predicted~\cite{alder_velocity_1967,alder_decay_1970,zwanzig_nonequilibrium_2001}. The simulation results were found to be in agreement with the predicted power-law decays only for Lennard-Jones fluids; instead a decay of $t^{-5/2}$ was extracted for water and a supercooled fluid~\cite{lesnicki_molecular_2016,straube_rapid_2020}. 
Here, we find that the long-time behavior of these time-dependent properties is significantly affected by finite-size effects arising from hydrodynamic interactions with periodic replicas, which masks the predicted long-time tails. 
Analytic corrections were previously developed in the stationary limit for the diffusion and friction coefficients~\cite{dunweg_molecular_1993-1,yeh_system-size_2004,simonnin_diffusion_2017}, based on the stationary Stokes equation.
Extending the calculations of D\"unweg et al.~\cite{dunweg_molecular_1993,dunweg_molecular_1993-1} and Yeh and Hummer~\cite{yeh_system-size_2004}, we derive a frequency-dependent finite-size correction allowing to retrieve the predicted asymptotic behavior from finite-size simulations. 
The method developed in this Letter is also applicable to other kinds of friction responses and more complex coarse-grained coordinates.

\begin{figure}
    \centering
    \includegraphics[width=0.5\textwidth]{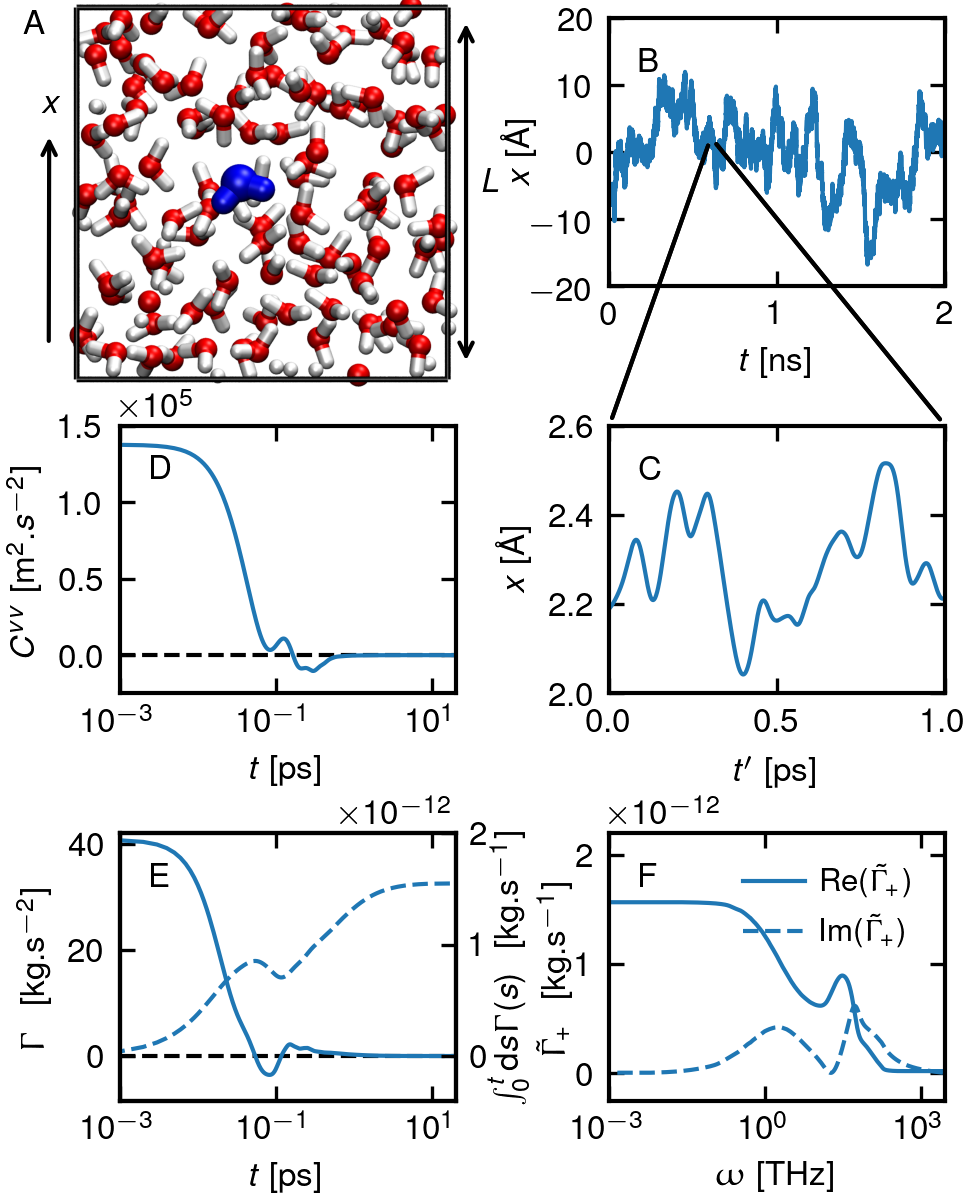}
    \caption{(A) Snapshot of a cubic simulation box of length $L=1.5$~nm filled with SPC/E water molecules. In blue a single water molecule is highlighted. Position trajectory $x$ of a single water molecule center of mass as a function of time $t$ on the nanosecond (B) and on the picosecond timescale (C). (D) Velocity autocorrelation function $C^{vv}$ averaged over all water molecules in the simulation box as a function of time. (E) Memory kernel $\Gamma$ (solid line) and integrated friction $\int_0^{t} {\rm d}s \Gamma(s)$ (dashed line). (F) Fourier transform of the memory kernel $\Tilde{\Gamma}_+$.
    }
    \label{fig:intro}
\end{figure}

In this study, we investigate SPC/E water~\cite{berendsen_missing_1987} as well as a Lennard-Jones (LJ) fluid with parameters corresponding to liquid argon~\cite{fachin2012}, for which results are shown in Appendix~\ref{app:lj}. In both cases, we simulate cubic boxes of length $L$ using 3D PBC for a range of box lengths $L$ from 1.5 to 5.0~nm (simulation details are provided in Appendix~\ref{app:sim}). Fig.~\ref{fig:intro}A shows a typical snapshot of the water simulation box, alongside a typical trajectory of the $x$ component of a single water molecule (tagged in blue) in Figs.~\ref{fig:intro}B-C, at different timescales. Panel C focuses on the picosecond timescale, which displays ballistic motion, while the nanosecond scale in panel B shows the Brownian diffusive regime. For longer times, the unwrapped water position diffuses away from its initial position.

We consider in this work the Mori GLE~\cite{mori_transport_1965} for the position of a particle of mass $m=k_BT/\langle v^2\rangle$ with velocity $\vec{v}$, given in the absence of a potential as
\begin{equation}\label{eq:moriGLE}
    m \dot{v_i}(t) = - \int_0^\infty {\rm d}s \, \Gamma_{ij}(t-s) v_j(s) + F_i^R(t) \,,
\end{equation}
where the random force $F^R$ has zero mean and is related to the memory kernel $\Gamma$ by the fluctuation-dissipation theorem $\langle F_i^R(t) F_i^R(0)\rangle = k_BT \Gamma_{ii}(t)$, with $i= x, y, z$. 
We introduce here a memory tensor $\Gamma_{ij}(t) = \delta_{ij} \Gamma(t)$, which by isotropy has no off-diagonal correlations.
To extract the memory kernel from simulation trajectories, we use a second-order Volterra iterative scheme~\cite{kowalik_memory-kernel_2019} (see Appendix~\ref{app:volterra}), which only depends on the VACF $C^{vv}$. 
Fig.~\ref{fig:intro} showcases the kernel extraction from a water simulation: panel D displays $C^{vv}$, panel E the memory kernel $\Gamma$ (solid line) as well as its running integral (dashed line). The integral of the memory kernel links the GLE formalism to the steady-state hydrodynamic picture with a static friction coefficient $\gamma = \int _0^\infty {\rm d}s \, \Gamma(s)$. Finally, panel F shows the Fourier transform (FT) of the memory kernel, which plays a key role in this Letter, as we derive the finite-size correction in frequency space. We take the FT of a function $f(\vec{r}, t)$ to be $\Tilde{f}(\vec{r}, \omega) = \int {\rm d}t \, {\rm e}^{i \omega t} f(\vec{r}, t)$ and consider for the memory kernel the single-sided FT $\Tilde{\Gamma}_+(\omega) = \int_0^\infty {\rm d}t {\rm e}^{i \omega t} \Gamma(t)$. The real part of $\Tilde{\Gamma}_+$ in Fig.~\ref{fig:intro}F plateaus for low frequencies and decays to zero for high frequencies, while the imaginary part vanishes both at low and high frequencies.

\begin{figure}
    \centering
    \includegraphics[width=0.5\textwidth]{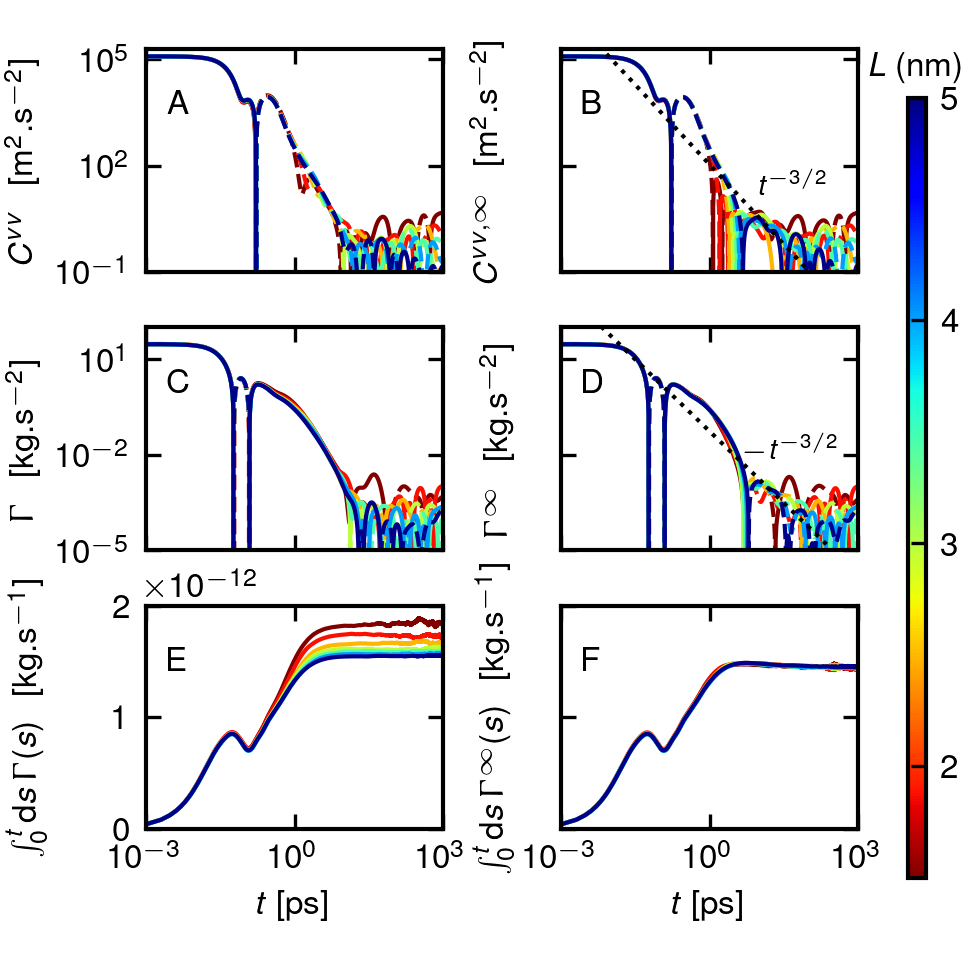}
    \caption{Velocity autocorrelation function $C^{vv}(t)$ (A-B), memory kernels $\Gamma(t)$ (C-D) and integrated friction $\int_0^{t} {\rm d}s \Gamma(s)$ (E-F) for the center of mass position of an SPC/E water molecule in water, for different box sizes $L \in [1.5, 5]$~nm, extracted directly from MD simulations (A, C, E) and corrected for finite size effects using Eq.~\ref{eq:cvv_corr} (B) and Eq.~\ref{eq:correction} (D, F). Dashed lines in log-log plots indicate negative values and the data is smoothed using a Gaussian filter in log space. The dotted black lines are power-law decays $t^{-3/2}$ as predicted by the long-time tail Eq.~\ref{eq:gamma_tail}.
    }
    \label{fig:corrected}
\end{figure}

From extensive molecular simulations, we extract memory kernels for different box sizes $L$ ranging from 1.5 to 5~nm. Figs.~\ref{fig:corrected}A, C, E show the extracted VACF, memory kernels and kernel integrals for water. These properties show little variations for short times, while the long-time behavior displays a significant box-size dependence. Note that this long-time regime is particularly susceptible to numerical noise, so that the $L$-dependence is most visible in the integral of the memory kernel in Fig.~\ref{fig:corrected}E, which plateaus at different friction coefficient values $\gamma$ depending on the box-size. The dependence of $\gamma$ on box size was investigated earlier~\cite{dunweg_molecular_1993,dunweg_molecular_1993-1,yeh_system-size_2004}, and we verify in Appendix~\ref{app:yeh-hummer} that $\gamma^{-1}$ is inversely proportional to $L$ with the expected proportionality constant~\cite{yeh_system-size_2004}. Most importantly, for the investigated box sizes we do not observe the long-time tail predicted by Eq.~\ref{eq:gamma_tail}, neither for the memory kernels nor for the VACF. 

In order to correct for the effect of PBC on the memory kernel, we start from the transient Stokes equation: the frequency-dependent velocity field $\Tilde{v}(\vec{r}, \omega)$ due to an external force $\Tilde{F}(\vec{r}, \omega)$ acting on the fluid is given by a convolution with the tensorial Green's function $G$. It can be separated into a transverse $G^T$ and a longitudinal $G^L$ contribution, given explicitly both in Fourier and real space in Ref.~\cite{erbas_viscous_2010} and in Appendix~\ref{app:erbas}. We only need the trace of the Green's functions for the calculation, which are given by
\begin{equation}\label{eq:trg}
    \frac{1}{3}{\rm Tr}[\Tilde{G}_{ij}^T(\vec{r}, \omega)] = \frac{{\rm e}^{-\alpha r}}{6 \pi \eta r} \quad \text{and} \quad
    \frac{1}{3}{\rm Tr}[\Tilde{G}_{ij}^L(\vec{r}, \omega)] = \frac{\lambda^2}{\alpha^2}\frac{{\rm e}^{-\lambda r}}{12 \pi \eta r}\,,
\end{equation}
where we introduced two characteristic lengths $\alpha^{-1}(\omega)$ and $\lambda^{-1}(\omega)$
\begin{equation}
    \alpha^2 = \frac{-i \omega \rho}{\eta} \quad \text{and} \quad 
    \lambda^2 = \frac{-i \omega \rho}{4\eta / 3 + \zeta + i\rho c^2 \ \omega} \,,
\end{equation}
with $\zeta$ the volume viscosity and $c$ the speed of sound. In the limit of an incompressible fluid, $c \to \infty$, one has $\lambda \to 0$ and thus the longitudinal contribution vanishes.

Let us now consider a cubic system of size $L$ with PBC, where we apply a point force at $\vec{r}=\vec{0}$. The force applied in the unit cell has infinitely many periodic images so that the total force field is expressed as $\Tilde{F}_i(\vec{r}, \omega) = \left[\left(\sum _{\vec{n}}\delta(\vec{r} + \vec{n}L)\right)-1/L^3\right]\Tilde{F}_i(\omega)$, where $\vec{n} = n_x \vec{e}_x + n_y \vec{e}_y + n_z \vec{e}_z$ is a lattice vector with $n_x, n_y, n_z$ integers and $\vec{e}_i$ the unit vectors in directions $x, y, z$. Note that we added a uniform background force to ensure momentum conservation~\cite{yeh_system-size_2004}. 
The force within the periodic images results in hydrodynamic interactions and induces a spurious velocity field contribution, which depends on the box size and can be written as a convolution of the tensor $G$ and the applied forces. For $\vec{r} = \vec{0}$ and using the Einstein summation convention, this gives
\begin{align}\label{eq:vPBCgen}
    \Delta\Tilde{v}_i^{\rm corr}(\omega)
    &= \int {\rm d}\vec{r}' \Tilde{G}_{ij}(\vec{r}', \omega) \left[ \left(\sum\limits_{\vec{n}, \vec{n} \neq 0} \delta(\vec{n}L - \vec{r}')\right) - \frac{1}{L^3} \right] \Tilde{F}_j(\omega) \nonumber \\
    &= \left[ \left(\sum\limits_{\vec{n}, \vec{n} \neq 0} \Tilde{G}_{ij}(\vec{n}L, \omega)\right) - \frac{1}{L^3} \int {\rm d}\vec{r}'\Tilde{G}_{ij}(\vec{r}', \omega)\right] \Tilde{F}_j(\omega)\,.
\end{align}
Indeed $\Delta\Tilde{v}_i^{\rm corr}$ results from the response to the point forces in the periodic images, excluding the central image, and from the background neutralising force. 
Introducing next the friction kernel extracted from MD simulations $\Gamma^{\rm MD}(t)$ and the one in the limit of an infinite system $\Gamma^{\infty}(t)$, and using the GLE Eq.~\ref{eq:moriGLE}, we obtain the relation between the velocity difference $\Delta \Tilde{v}^{\rm corr}$ and the friction force exerted by the fluid on the tagged particle as $\Delta \Tilde{v}_i^{\rm corr}(\omega) = ([\Tilde{\Gamma}^{\rm MD}_{+, ij}(\omega)]^{-1} - [\Tilde{\Gamma}_{+, ij}^{\infty}(\omega)]^{-1} )\Tilde{F}_j(\omega)$. 
Combining this with Eq.~\ref{eq:vPBCgen} the force $\Tilde{F}_j(\omega)$ drops out. After taking the trace we obtain
\begin{equation}
    [\Tilde{\Gamma}_{+}^{\infty}(\omega)]^{-1} = [\Tilde{\Gamma}_{+}^{\rm MD}(\omega)]^{-1} - \Delta \Tilde{G}^{\rm corr}(\omega) \,, \label{eq:correction}
\end{equation}
where we introduced
\begin{equation}\label{eq:gcorr}
    \Delta \Tilde{G}^{\rm corr}(\omega) = \left[\sum\limits_{\vec{n}, \vec{n} \neq 0} \frac{1}{3} {\rm Tr}[\Tilde{G}_{ij}(\vec{n}L, \omega)] \right] - \frac{1}{3L^3} \int {\rm d}\vec{r}' {\rm Tr}[\Tilde{G}_{ij}(\vec{r}', \omega)] \,.
\end{equation}
This is the main result of this Letter, which gives an explicit expression for the effect of PBC on the memory kernel and allows to calculate the infinite box size friction kernel $\Gamma^{\infty}(t)$ from the simulated finite box size kernel $\Gamma^{\rm MD}(t)$.
This frequency-dependent correction can readily be applied to the velocity autocorrelation function $C^{vv}$ (see Appendix~\ref{app:cvv}) and yields
\begin{equation}
    \Tilde{C}_{+}^{vv, \infty}(\omega) = \frac{\Tilde{C}_{+}^{vv, \rm MD}(\omega)}{1 + (k_BT)^{-1}\Tilde{C}_{+}^{vv, \rm MD}(\omega)\Tilde{\Gamma}_+^{\infty}(\omega)\Tilde{\Gamma}_+^{\rm MD}(\omega)\Delta \Tilde{G}^{\rm corr}(\omega)} \,.\label{eq:cvv_corr}
\end{equation}
The mean-squared displacement follows by double integration.
We further provide explicit forms to compute the transverse contribution to the correction $\Delta \Tilde{G}^{T, \rm corr}$. Using Eq.~\ref{eq:trg}, we explicitly write the transverse correction defined by Eq.~\ref{eq:gcorr} as
\begin{align}
    \Delta \Tilde{G}^{T, \rm corr}(\omega) 
    &= \frac{1}{6 \pi \eta } \left[ \sum \limits_{\vec{n}, \vec{n}\neq 0} \frac{{\rm e}^{-\alpha \vert \vec{n} \vert L}}{\vert \vec{n} \vert L} \right]
    - \frac{2}{3 \eta \alpha^2 L^3} \label{eq:gcorr_r_fin}\,.
\end{align}
For large $\alpha$, the real space sum in Eq.~\ref{eq:gcorr_r_fin} converges quickly. To also cover the low frequency regime, \textit{i.e.} for small $\alpha$, we transform Eq.~\ref{eq:gcorr} using an Ewald summation (for explicit expressions, comparison and convergence studies, see Appendix~\ref{app:ewald}). For $\omega \to 0$, we retrieve Yeh and Hummer's zero-frequency correction~\cite{yeh_system-size_2004} as expected. Equivalent results are straightforwardly derived for the longitudinal contribution (see Appendix~\ref{app:long}). In the following, we show results for the hydrodynamic correction with both transverse and longitudinal contributions computed with the Ewald expression.

\begin{figure}[hbt!]
    \centering
    \includegraphics[width=0.5\textwidth]{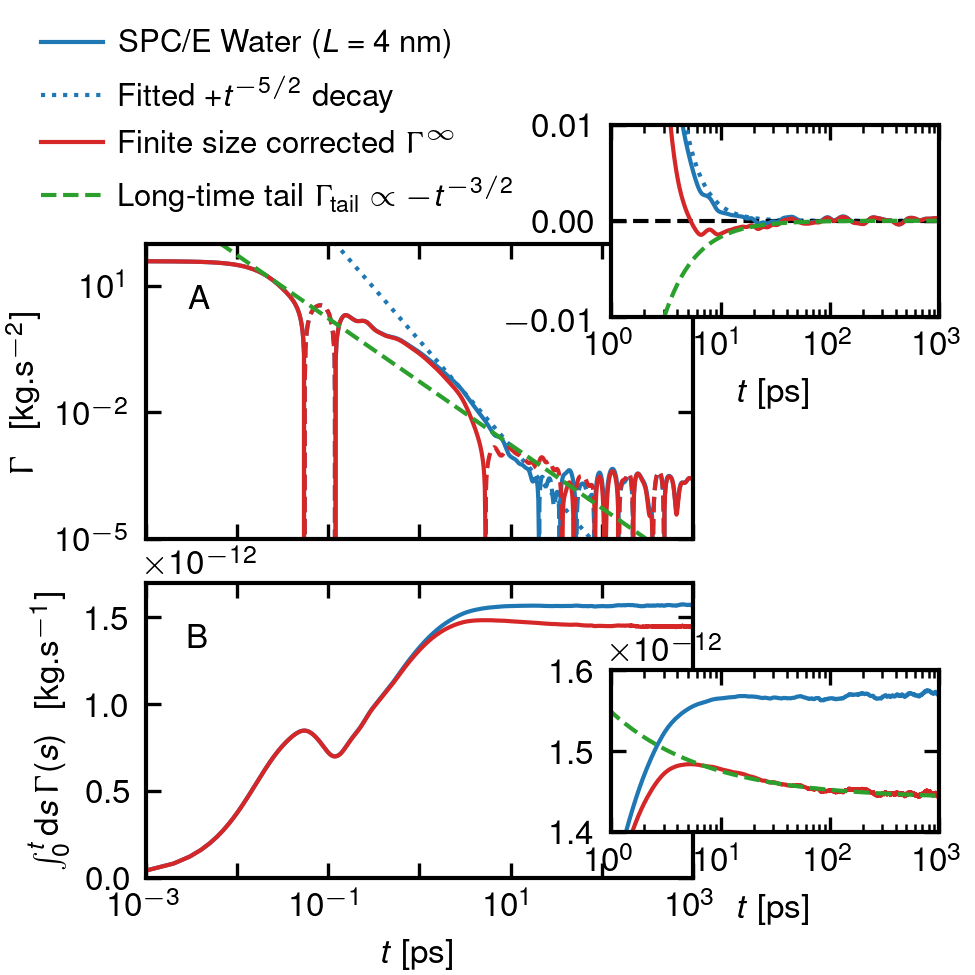}
    \caption{Memory kernel $\Gamma(t)$ (A) and integrated friction $\int_0^{t} {\rm d}s \Gamma(s)$ (B) for the center of mass position of an SPC/E water molecule in water, extracted from MD simulations (blue line) and corrected for finite-size effects using Eq.~\ref{eq:correction} (red line). Dashed lines in log-log plots indicate negative values and the data is smoothed using a Gaussian filter in log space. We show the predicted hydrodynamic long-time tail $\Gamma_{\rm tail}(t)$ in Eq.~\ref{eq:gamma_tail} (green dashed line), computed using the corrected value of the friction coefficient $\gamma^\infty$ (see Appendix~\ref{app:yeh-hummer}) in $\gamma$ and $D = k_B T/\gamma$, as well as a power-law fit $y=0.5t^{-5/2}$ to the extracted kernel (blue dotted line) suggested by Ref.~\cite{straube_rapid_2020}.
    }
    \label{fig:hydro}
\end{figure}

Figs.~\ref{fig:corrected}B, D, F present the corrected VACF $C^{vv, \infty}$, memory kernels $\Gamma^\infty$ and friction integrals. All curves from different box sizes fall onto a master curve, validating our method to correct these time-dependent response functions for finite-size effects. Additionally, we show in Appendix~\ref{app:fdep} that using frequency-dependent viscosity spectra $\Tilde{\eta}(\omega)$ and $\Tilde{\zeta}(\omega)$ extracted from MD simulations results in an even better superposition of the different curves, pointing to a more accurate finite-size correction. 
Strikingly, our correction modifies the long-time power-law decay of the VACF and the memory kernel.
In Fig.~\ref{fig:hydro}, we compare the extracted $\Gamma^{\rm MD}$ for a box length $L = 4$~nm (blue line) and the corrected $\Gamma^\infty$ (red line) with the predicted hydrodynamic long-time tail Eq.~\ref{eq:gamma_tail}. For long times, $\Gamma^{\rm MD}(t)$ is positive and decays as $+t^{-5/2}$~\cite{straube_rapid_2020} (blue dotted line). However, this is only a spurious decay due to the PBC: the finite-size correction modifies the kernels at times longer than 1~ps and as a consequence reveals the negative long-time tail in Eq.~\ref{eq:gamma_tail} proportional to $-t^{-3/2}$ (green dashed line) in the kernels, which results in a decay as $-t^{-1/2}$ of the friction integral for times larger than $\sim$ 1~ps. The agreement with the long-time tail Eq.~\ref{eq:gamma_tail} is excellent. We draw similar conclusions for the VACF and its long-time tail, as shown in Fig.~\ref{fig:corrected} (dotted lines) and in Appendix~\ref{app:cvv}. The results for a LJ particle in a LJ fluid are given in Appendix~\ref{app:lj} and support our conclusions. 
This demonstrates the importance of taking into account hydrodynamic interactions due to PBC and correcting time-dependent quantities such as the memory kernel and the VACF when investigating hydrodynamics and long-time behaviors. This correction further allows to reduce the computational effort and memory (in terabytes) of such studies, and to explore even longer-time behaviors.

\begin{figure}
    \centering
    \includegraphics[width=0.5\textwidth]{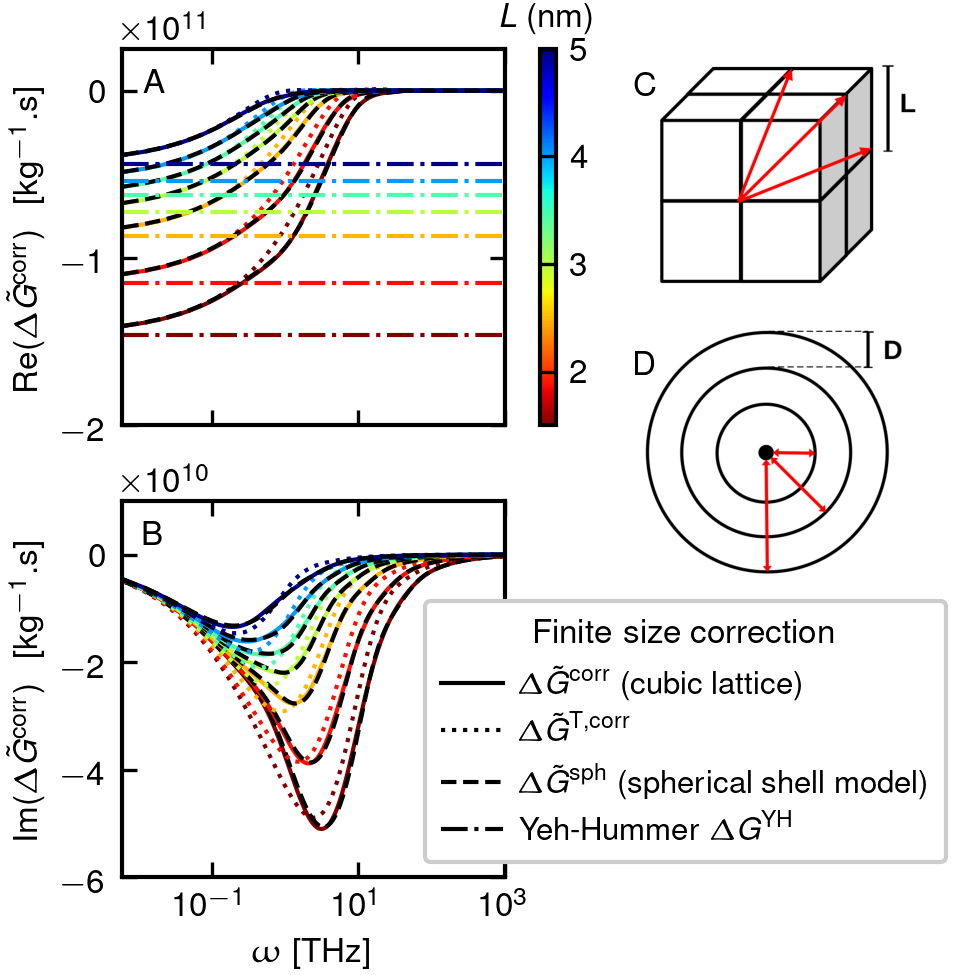}
    \caption{Real (A) and imaginary (B) parts of the frequency-dependent finite size correction $\Delta \Tilde{G}^{\rm corr}(\omega)$, for the viscosity and density of SPC/E water, for a range of box sizes $L \in [1.5, 5]$~nm shown by the colorbar. We show the total correction including transverse and longitudinal contributions $\Delta \Tilde{G}^{\rm corr}$ in Eq.~\ref{eq:gcorr} (solid line), the transverse contribution to the correction $\Delta \Tilde{G}^{\rm T, corr}$ (Eq.~\ref{eq:gcorr_k_fin} in Appendix~\ref{app:ewald}, dotted lines), as well as the spherical shell model $\Delta \Tilde{G}^{\rm sph}$ given in Eq.~\ref{eq:fit} (dashed black lines).     For reference, we give the correction derived by Yeh and Hummer in the zero-frequency limit $\Delta G^{\rm YH} \approx -2.837297 / (6\pi\eta L)$~\cite{yeh_system-size_2004} (horizontal dash-dotted lines).
    }
    \label{fig:correction}
\end{figure}

Finally, to simplify the use of our frequency-dependent finite-size correction scheme, we introduce an exactly solvable model consisting of concentric spherical shells at a radial separation $D$, at which constant surface force densities act, as schematized in Fig.~\ref{fig:correction}D, instead of the cubic periodic lattice considered up to now and drawn in Fig.~\ref{fig:correction}C. This spherical shell model yields a simple functional form
\begin{equation}
    \Delta \Tilde{G}^{\rm sph}(\omega) 
    = \frac{D^2}{3\eta L^3} \left[ 2 f(\alpha^{-1})
    + \frac{\lambda^2}{\alpha^2}f(\lambda^{-1}) 
    - \frac{3}{D^2\alpha^2} \right]
     \,,\label{eq:fit}
\end{equation}
where $f(x) = {\rm e}^{-m_0D/x}[e^{D/x}(1+m_0)-m_0]/({e}^{D/x}-1)^2$ comes from the sum over periodic spheres. The derivation of this expression is given in Appendix~\ref{app:spherical} and includes both transverse and longitudinal contributions.
We fix the separation $D$ so that the zero-frequency limit equals the Yeh-Hummer expression $D^2=3\xi L^2 / [\pi (6m_0^2 + 6m_0 + 1)]$, with $\xi = 2.837297$, and we fit the parameter $m_0$ to the numerically determined correction $\Delta\Tilde{G}^{\rm corr}$, yielding $m_0 = 0.387$. Fig.~\ref{fig:correction} shows the comparison of the real part (panel A) and imaginary part (panel B) of the different corrections derived in this Letter. The agreement of $\Delta\Tilde{G}^{\rm sph}$ with $\Delta \Tilde{G}^{\rm corr}$ is excellent, so that $\Delta\Tilde{G}^{\rm sph}$ can safely be used in practical applications. Interestingly, the transverse contribution is the major part of $\Delta\Tilde{G}^{\rm corr}$, while the longitudinal part is almost negligible, \textit{i.e.} using only the transverse part of $\Delta \Tilde{G}^{\rm corr}$ is a good approximation. As expected, the real part of the correction retrieves Yeh and Hummer's for zero frequency (horizontal dash-dotted lines in Fig.~\ref{fig:correction}A), and the correction increases in magnitude with $1/L$. Moreover, there is a shift towards higher frequencies of the main features of the correction for smaller box lengths $L$, suggesting that the smaller the box size, the shorter the timescales influenced by hydrodynamic interactions.  

The frequency-dependent finite-size correction scheme developed in this Letter retrieves long-time dynamics, such as the long-time tails predicted by hydrodynamics, from simulations of relatively small systems, which is helpful for MD simulations of aqueous systems and important to encode the correct long-time dynamics for example in coarse-grained molecular simulations~\cite{klippenstein_introducing_2021}. This work opens the way to the treatment of more complex systems and observables~\cite{bocquet_friction_1997}, but could also be extended to other time-dependent transport properties such as electrophoresis, diffusiophoresis or thermal conductivity.

\begin{acknowledgments}
We acknowledge support by the ERC Advanced Grant No. 835117 NoMaMemo and by the Deutsche Forschungsgemeinschaft (DFG) via the project SFB 1449-431232613-A02. We gratefully acknowledge computing time on the HPC clusters at the Physics department and ZEDAT, FU Berlin.
\end{acknowledgments}


%

\clearpage

\appendix
\section{Simulation details}\label{app:sim}
For the water systems, we prepare 7 systems at $T=300$~K using the rigid SPC/E water model~\cite{berendsen_missing_1987}, with initial box lengths $L = 1.5$, 2.0, 2.5, 3.0, 3.5, 4.0 and 5.0~nm, with 109, 221, 510, 884, 1378, 2165 and 4055 water molecules, respectively. Simulations are run using the LAMMPS molecular simulation software~\cite{LAMMPS}.

For the Lennard-Jones fluid, we simulate 7 systems at $T=92$~K with initial box lengths $L = 2.2$, 2.5, 3.0, 3.5, 4.0, 4.5, 5.0~nm and 217, 343, 513, 1000, 1331, 1764 and 2745 LJ particles, respectively, using the GROMACS simulation package~\cite{abraham_gromacs_2015}. For all particles we took the Lennard-Jones parameters of argon of the GROMOS53a6 force field~\cite{fachin2012} ($\sigma = 3.410$~\AA, $\epsilon = 0.996$~kJ.mol$^{-1}$ and a cutoff radius of $2.5\sigma$). Using LJ units, the systems are at $T^* = 0.77$ and $P^* = 0.04$ corresponding to the liquid phase~\cite{vrabec_comprehensive_2006,ahmed_effect_2010}.

All systems are then equilibrated in the NPT ensemble (with $P = 1$~bar for water and $P=17$~bar for the LJ systems) for at least 500~ps (the final box length $L$ is only weakly modified), followed by a production run in the NVT ensemble for 10~ns for water and 20~ns for the LJ systems. The equations of motion are solved using the velocity Verlet algorithm using a timestep of 1~fs for water and 2~fs for the LJ fluid. For the LJ fluid, we use a velocity rescale thermostat \cite{bussi2007}, while for water, we use a Nose-Hoover thermostat with time constant 500~fs and constrain the geometry of water molecules using the RATTLE algorithm. 

The $x$, $y$ and $z$ coordinates of the center of mass of all water molecules and of all LJ particles are printed out at each timestep. 
The velocity autocorrelation function is then computed separately for each component using the Wiener-Khinchin theorem and the averaged velocity autocorrelation function over all three components and over all particles is used to compute the memory kernel. 

For SPC/E water, we obtain the density $\rho = 994$~kg.m$^{-3}$, the shear viscosity $\eta = 0.697 \cdot 10^{-3}$~Pa.s, the volume viscosity $\zeta = 1.73 \cdot 10^{-3}$~Pa.s and the sound velocity $c=1510$~m.s$^{-1}$~\cite{sedlmeier_chargemass_2014} (see Appendix~\ref{app:green_kubo} and \ref{app:fitting_procedure}).
For the LJ fluid, we find $\rho = 1370$~kg.m$^{-3}$, $\eta = 0.23 \cdot 10^{-3}$~Pa.s, $\zeta = 0.59 \cdot 10^{-3}$~Pa.s and $c = 869$~m.s$^{-1}$ obtained as $c^2 = {\rm d}P/{\rm d}\rho = \sqrt{\xi/\rho}$, with $\xi$ the bulk modulus~\cite{VANDAEL1966611,haynes2016crc}.
The kinematic viscosity $\eta / \rho$ is thus $7.01\cdot 10^{-7}$~m$^2$.s$^{-1}$ for SPC/E water and $1.68\cdot 10^{-7}$~m$^2$.s$^{-1}$ for the LJ fluid. Using the asymptotic result for the friction coefficient $\gamma$ from Appendix~\ref{app:yeh-hummer}, we obtain the diffusion coefficients as $D = k_B T / \gamma$, yielding $2.87\cdot 10^{-9}$~m$^2$.s$^{-1}$ for SPC/E water and $2.51\cdot 10^{-9}$~m$^2$.s$^{-1}$ for the LJ fluid. The diffusion coefficients are therefore 2 orders of magnitude smaller than the kinematic viscosities for both our systems, showing that the first term in Eqs.~\ref{eq:gamma_tail} and \ref{eq:cvv_tail} is negligible.

\section{Results for a Lennard-Jones fluid}\label{app:lj}
\begin{figure}[hbt!]
    \centering
    \includegraphics[width=0.5\textwidth]{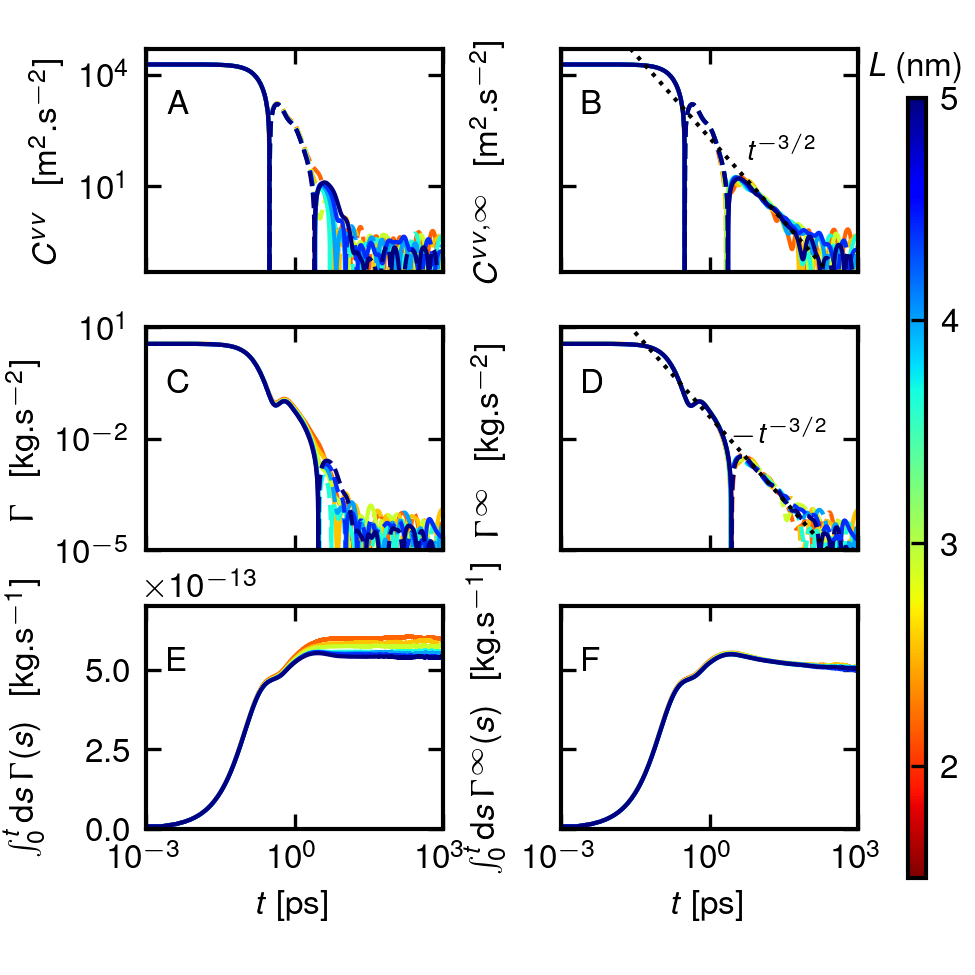}
    \caption{Finite size correction for the LJ fluid. Velocity autocorrelation function $C^{vv}(t)$ (A-B), memory kernels $\Gamma(t)$ (C-D) and integrated friction $\int_0^{t} {\rm d}s \Gamma(s)$ (E-F) extracted directly from MD simulations (A, C, E) and corrected for finite size effects using Eq.~\ref{eq:cvv_corr} (B) and Eq.~\ref{eq:correction} (D, F). Data is shown for the position of a LJ particle in a LJ fluid, for different box sizes $L \in [2.1, 5]$~nm shown in the colorbar. Dashed lines in log-log plots indicate negative values and the data is smoothed using a Gaussian filter in log space. The dotted black lines are power-law decays $t^{-3/2}$ as predicted by the long-time tail in Eq.~\ref{eq:gamma_tail}.}
    \label{fig:corrected_lj}
\end{figure}

\begin{figure}[hbt!]
    \centering
    \includegraphics[width=0.5\textwidth]{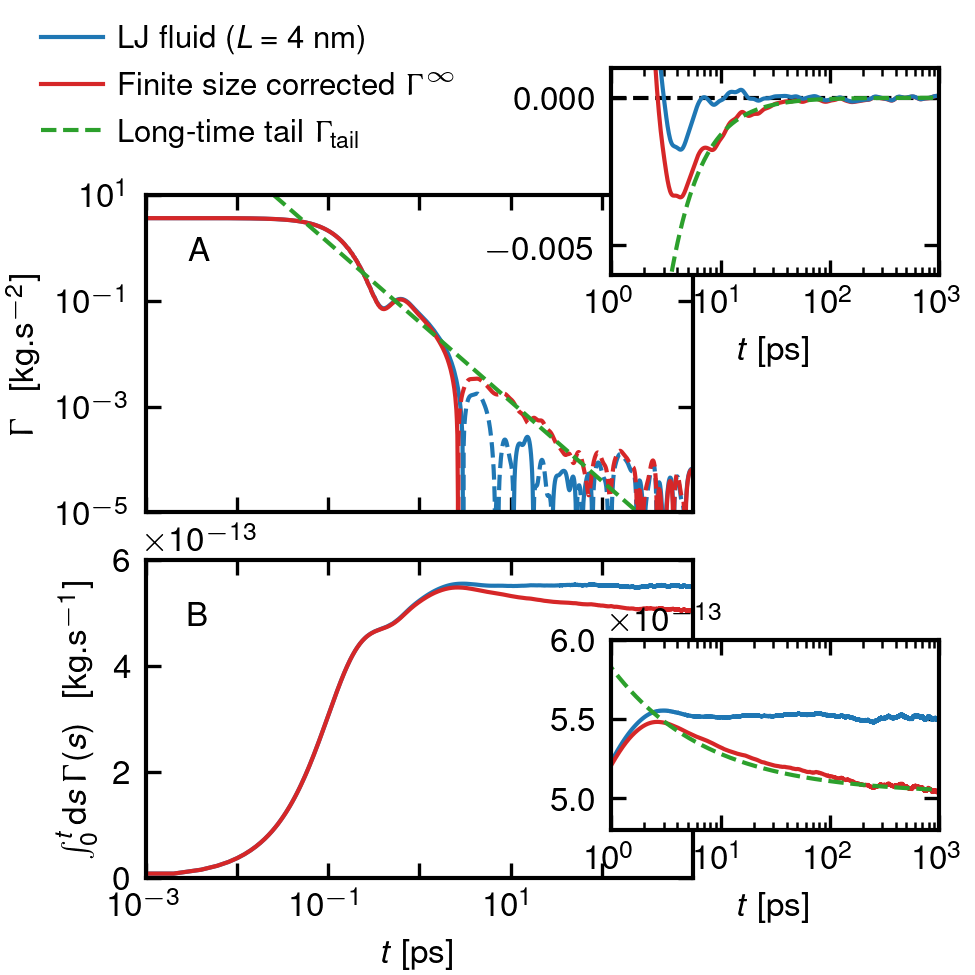}
    \caption{Hydrodynamic long-time tail for the LJ fluid. Memory kernel $\Gamma$ (A) and integrated friction $\int_0^{t} {\rm d}s \Gamma(s)$ (B) for the position of a LJ particle in a LJ fluid. Dashed lines in log-log plots indicate negative values and the data is smoothed using a Gaussian filter in log space. We show kernels extracted directly from MD simulations in blue and kernels corrected for finite-size effects using Eq.~\ref{eq:correction} in red. The results are compared with the predicted hydrodynamic long-time tail $\Gamma_{\rm tail}$ in Eq.~\ref{eq:gamma_tail} (green dashed line).}
    \label{fig:hydro_lj}
\end{figure}

We show here equivalent results as those given for a water molecule in the main text for the position of a LJ particle in a LJ fluid. The finite size correction is given in Fig.~\ref{fig:corrected_lj} and the hydrodynamic long-time tail in Fig.~\ref{fig:hydro_lj}.

\section{Second-order Volterra scheme}\label{app:volterra}
The extraction of memory kernels from MD simulations is done using a second order Volterra scheme, introduced earlier in the literature~\cite{kowalik_memory-kernel_2019}. Starting from the GLE in Eq.~\ref{eq:moriGLE}, using that $\langle F^R(t) \dot{x}(0)\rangle = 0$, we derive
\begin{align}
    m \frac{{\rm d} C^{vv}(t)}{{\rm d}t} &= - \int_0^t {\rm d}s \Gamma(s) C^{vv}_{t-s} \,, \label{eq:volt1}
\end{align}
where $C^{vv}(t) = \langle \dot{x}(t) \dot{x}(0) \rangle$ is the velocity autocorrelation function. Integrating Eq.~\ref{eq:volt1} yields
\begin{align}
    m\int_0^t {\rm d}t' \frac{{\rm d} C^{vv}(t')}{{\rm d}t'} &= - \int_0^t {\rm d}t' \int _0^{t'} {\rm d}s \, \Gamma(s) C^{vv}(t'-s)\\
    m [C^{vv}(t) - C^{vv}(0)] & =  - \int_0^t {\rm d}t' \int _0^{t'} {\rm d}s \, \Gamma(t'-s) C^{vv}(s) \nonumber\\
    & =  - \int_0^t {\rm d}s \int _s^t {\rm d}t' \,  \Gamma(t'-s) C^{vv}(s) \nonumber\\
    & =  - \int_0^t {\rm d}s \, C^{vv}(s) K(t-s) \nonumber\\
    m [C^{vv}(t) - C^{vv}(0)] & = - \int_0^t {\rm d}s \, K(s) C^{vv}(t-s) \,, \label{eq:volt2}
\end{align}
where we introduce the memory kernel integral $K(t) = \int_0^t {\rm d}s \, \Gamma(s)$. Using that $K(0) = 0$, we discretize Eq.~\ref{eq:volt2} and obtain the following iterative extraction scheme
\begin{equation}
    K_i = \frac{2}{\delta t C_0^{vv}} \left(m C_0^{vv} - m C_i^{vv} - \delta t \sum\limits_{j=1}^{i-1} C_j^{vv} K_{i-j} \right)\,
\end{equation}
where $K_i$ is the discrete integral of the memory function $K(t)$ and $C_i^{vv}$ is the discrete velocity autocorrelation function at time $t=i\delta t$, with $\delta t$ the timestep.

\section{Yeh-Hummer zero-frequency finite-size correction}\label{app:yeh-hummer}
\begin{figure}[hbt!]
    \centering
    \includegraphics[width=0.5\textwidth]{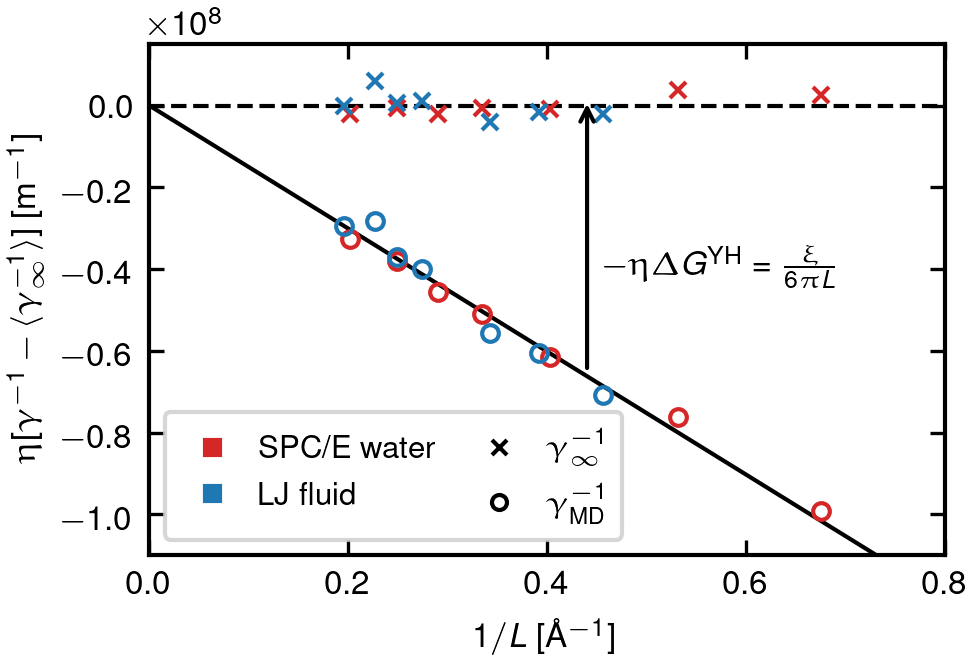}
    \caption{Coefficient $\eta (\gamma^{-1} - \langle \gamma^{-1}_\infty \rangle)$ as a function of the inverse box size $1/L$. Data is shown for SPC/E water (red symbols) and for a LJ fluid (blue symbols). Circles correspond to using friction coefficients extracted from MD simulations (taken as $\gamma_{\rm MD} = {\rm Re}[\Gamma_+^{\rm MD}(\omega = 0)]$), while crosses are the corrected values $\gamma_\infty$ using Eqs.~\ref{eq:yh1} and \ref{eq:yh2}. As a guide to the eye, the solid black line indicates the predicted linear inverse dependence on the box size $L$.}
    \label{fig:yh}
\end{figure}

The hydrodynamic correction for the static friction coefficient (related to the diffusion coefficient by $D = k_BT / \gamma$) was previously given by Yeh and Hummer~\cite{yeh_system-size_2004} as
\begin{equation}
    \gamma^{-1}_\infty = \gamma^{-1}_{\rm MD} - \Delta G^{\rm YH} \,.\label{eq:yh1}
\end{equation}
The authors derived the correction term as an Ewald sum that reduces to the expression
\begin{equation}
    \Delta G^{\rm YH} = - \frac{\xi}{6 \pi \eta L} \,, \label{eq:yh2}
\end{equation}
with the numerically determined constant $\xi = 2.837297$.
In Fig.~\ref{fig:yh}, we show how this correction reproduces quantitatively our simulation results in the case of SPC/E water (red symbols) and of a LJ fluid (blue symbols).

\section{Tensorial Green's function for an infinite system}\label{app:erbas}
We reproduce here the results of Ref.~\cite{erbas_viscous_2010} for the solution of the transient Stokes equation
\begin{equation}\label{eq:stokes}
    \rho \frac{\partial v_i}{\partial t} = F_i 
    - \nabla_i P 
    + \left( \frac{\eta}{3} + \zeta \right) \nabla_i \nabla_k v_k
    + \eta \nabla_k \nabla_k v_i \,,
\end{equation}
with $\vec{v}(\vec{r}, t)$ the velocity field at position $\vec{r}$ and time $t$, $\vec{F}(\vec{r}, t)$ the external force acting on the fluid at $\vec{r}$ and $P(\vec{r}, t)$ the pressure.
In the main text, we define the time-FT as $\Tilde{f}(\vec{r}, \omega) = \int {\rm d}t\, {\rm e}^{i \omega t} f(\vec{r}, t)$. Here, we additionally take the time- and space-FT of a function $f(\vec{r}, t)$ to be $\Hat{f}(\vec{k}, \omega) = \int {\rm d}t \int _{\mathbb{R}^3} {\rm d}^3 r \, {\rm e}^{-i \vec{k}\cdot\vec{r} + i \omega t} f(\vec{r}, t)$. 

The corresponding tensorial Green's function in Fourier space is found by separating the velocity field into a transverse and a longitudinal contribution $\Hat{v}_i = \Hat{v}^T_i + \Hat{v}^L_i$, such that $k_i \Hat{v}^T_i = 0$ and $k_i \Hat{v}^L_i = k_i \Hat{v}_i$. 
The Green's functions, defined as $\Hat{v}^T_i = \Hat{G}^T_{ij} \Hat{F}_j$ and $\Hat{v}^L_i = \Hat{G}^L_{ij} \Hat{F}_j$, are then given by
\begin{align}
    \Hat{G}^T_{ij}(\vec{k}, \omega) &= \frac{ \delta_{ij} - k_i k_j/k^2}{\eta (k^2 + \alpha^2)} \,,\quad \alpha^2 = \frac{-i \omega \rho}{\eta}\label{eq:ftgt} \\
    \Hat{G}^L_{ij}(\vec{k}, \omega) &= \frac{ k_i k_j \lambda^2}{\eta \alpha^2 k^2 (k^2 + \lambda^2)} \,,\quad \lambda^2 = \frac{-i \omega \rho}{4\eta / 3 + \zeta + i\rho c^2 \ \omega} \,.\label{eq:ftgl}
\end{align}
In the limit of an incompressible fluid, the speed of sound $c \to \infty$ and $\lambda \to 0$ and the longitudinal contribution vanishes. The tensors in real space are given by back Fourier transform as
\begin{align}
    \Tilde{G}^T_{ij}(\vec{r}, \omega) &= \frac{1}{4\pi\eta\alpha^2r^3}
    \Big[ \delta_{ij} ((1 + r\alpha + r^2 \alpha^2) {\rm e}^{-r\alpha} - 1)  \nonumber \\
    &\qquad + \frac{3r_i r_j}{r^2}(1 - (1 + r\alpha + r^2 \alpha^2 / 3) {\rm e}^{-r\alpha})\Big] \label{eq:rgt}\\
    \Tilde{G}^L_{ij}(\vec{r}, \omega) &= \frac{1}{4\pi\eta\alpha^2r^3}
    \Big[ \delta_{ij} (1 - (1 + r\lambda) {\rm e}^{-r\lambda})  \nonumber\\
    &\qquad - \frac{3r_i r_j}{r^2}(1 - (1 + r\lambda + r^2 \lambda^2 / 3) {\rm e}^{-r\lambda})\Big] \,. \label{eq:rgl}
\end{align}

\section{Finite-size correction of the velocity autocorrelation function}\label{app:cvv}
\begin{figure}[hbt!]
    \centering
    \includegraphics[width=0.5\textwidth]{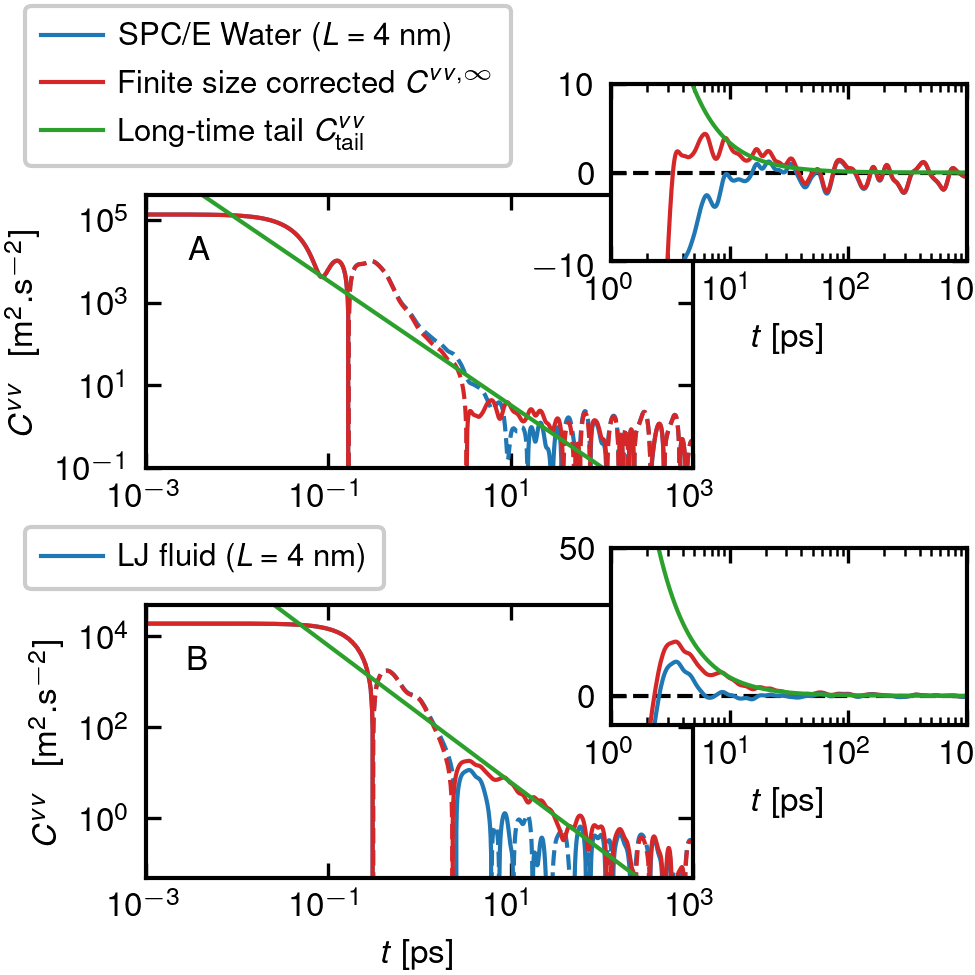}
    \caption{Hydrodynamic long-time tail of the velocity autocorrelation function $C^{vv}$ for the center of mass position of an SPC/E water molecule in water (A) and for a LJ particle in a LJ fluid (B). Dashed lines in log-log plots indicate negative values and the data is smoothed using a Gaussian filter in log space. We show the VACF extracted directly from MD simulations in blue and the VACF corrected for finite-size effects using Eq.~\ref{eq:cvv_corr2} (Eq.~\ref{eq:cvv_corr} of the main text) in red. The results are compared with the predicted hydrodynamic long-time tail $C^{vv}_{\rm tail}$ in Eq.~\ref{eq:cvv_tail} (green solid line), computed using the corrected value $\gamma^\infty$ in $D = k_B T / \gamma$ (see Appendix~\ref{app:yeh-hummer}).}
    \label{fig:cvv}
\end{figure}

Using the GLE and the fluctuation-dissipation theorem~\cite{kowalik_memory-kernel_2019}, we write the relation between the velocity autocorrelation function (VACF) single-sided Fourier transform $\Tilde{C}^{vv}_+(\omega) = \int_0^\infty {\rm d}t {\rm e}^{i \omega t} C^{vv}(t)$ and the memory kernel as
\begin{equation}
    \Tilde{C}_{+}^{vv}(\omega) = \frac{k_BT}{-i\omega m + \Tilde{\Gamma}_+(\omega)} \,.
\end{equation}
Using the finite-size correction in Eq.~\ref{eq:correction}, we relate the VACF extracted from MD,
\begin{equation}
    \Tilde{C}_{+}^{vv, \rm MD}(\omega) = \frac{k_BT}{-i\omega m + \Tilde{\Gamma}_+^{\rm MD}(\omega)}\,,
\end{equation}
to the infinite system limit $C_{vv}^{\infty}$ as
\begin{align}
    \Tilde{C}_{+}^{vv, \infty}(\omega) &= \frac{k_BT}{-i\omega m + \Tilde{\Gamma}_+^{\infty}(\omega)} \\
    \Tilde{C}_{+}^{vv, \infty}(\omega) &= \frac{\Tilde{C}_{+}^{vv, \rm MD}(\omega)}{1 + (k_BT)^{-1}\Tilde{C}_{+}^{vv, \rm MD}(\omega)\Tilde{\Gamma}_+^{\infty}(\omega)\Tilde{\Gamma}_+^{\rm MD}(\omega)\Delta \Tilde{G}^{\rm corr}(\omega)} \,.\label{eq:cvv_corr2}
\end{align}
Eq.~\ref{eq:cvv_corr2} is the equivalent of Eq.~\ref{eq:correction} for the memory kernel but for the VACF instead. Fig.~\ref{fig:corrected}A-B shows the correction of the VACF using Eq.~\ref{eq:cvv_corr2} and we show in Fig.~\ref{fig:cvv} the long-time tail of the VACF given as~\cite{alder_velocity_1967,alder_decay_1970,zwanzig_nonequilibrium_2001,corngold_behavior_1972,lesnicki_molecular_2016}
\begin{equation}
    C^{vv}_{\rm tail}(t) = \frac{2k_BT}{3 \rho} \left[4 \pi \left(D + \frac{\eta}{\rho}\right)t\right]^{-3/2}\,. \label{eq:cvv_tail}
\end{equation}

\section{Alternative expression for the finite-size correction using the Ewald summation}\label{app:ewald}
\begin{figure}
    \centering
    \includegraphics[width=0.5\textwidth]{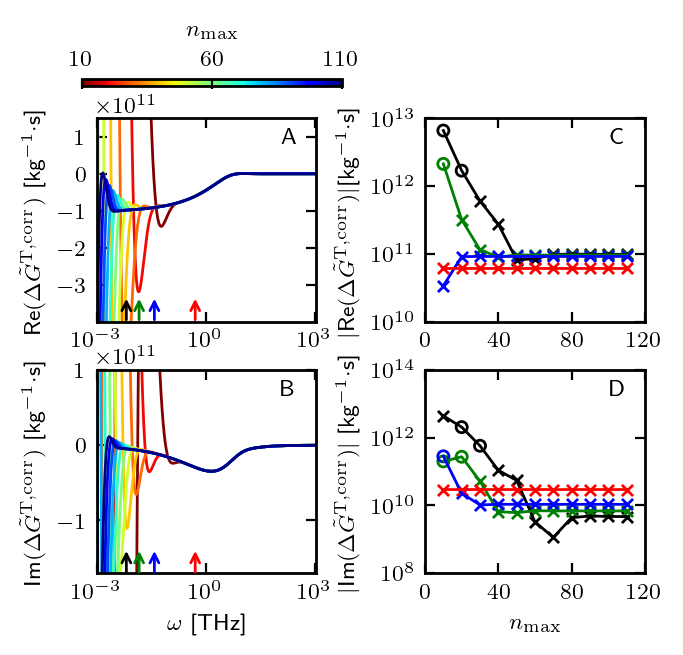}
    \caption{Test of the convergence of the real space summation for the transverse part Eq.~\ref{eq:gcorr_r_fin}. Frequency-dependent real (A) and imaginary (B) part of Eq.~\ref{eq:gcorr_r_fin} for different truncation indices $n_{\rm{max}} = \llbracket 10,20\dots110\rrbracket$, using the parameters of SPC/E $\rho = 994$~kg.m$^{-3}$, $\eta = 0.697 \cdot 10^{-3}$~Pa.s and a box size $L=2$~nm. Panels (C) and (D) show the correction in Eq.~\ref{eq:gcorr_r_fin} at $\omega=5\cdot10^{-3},~1\cdot10^{-2},~4\cdot10^{-2},~0.5$~THz (indicated by the arrows in panels A and B) as a function of the truncation index $n_{\rm{max}}$. Crosses and circle markers indicate respectively positive and negative values.}
    \label{fig:real_conv}
\end{figure}

The real space sum in Eq.~\ref{eq:gcorr_r_fin} is expected to converge well for high frequencies but converges slowly for small frequencies. This is shown numerically in Fig.~\ref{fig:real_conv} for a given box size $L=1.5$~nm, for which the sum reaches convergence for $n_{\rm max} > 80$.
We thus derive here an Ewald summation for the correction $\Delta G^{\rm corr}$, which converges faster, especially for small frequencies.
In the main text, we define the time-FT as $\Tilde{f}(\vec{r}, \omega) = \int {\rm d}t\, {\rm e}^{i \omega t} f(\vec{r}, t)$. Here, we additionally take the time- and space-FT of a function $f(\vec{r}, t)$ to be $\Hat{f}(\vec{k}, \omega) = \int {\rm d}t \int _{\mathbb{R}^3} {\rm d}^3 r \, {\rm e}^{-i \vec{k}\cdot\vec{r} + i \omega t} f(\vec{r}, t)$.

We first propose an alternative expression of Eq.~\ref{eq:gcorr} in Fourier space. For this, we use the Poisson summation formula 
\begin{equation}
    \sum\limits_{\vec{n}} \Tilde{G}_{ij}(\vec{r} + \vec{n}L, \omega) = \frac{1}{L^3} \sum\limits_{\vec{k}} \Hat{G}_{ij}(\vec{k}, \omega) {\rm e}^{i\vec{k}\cdot\vec{r}}\,,  \label{eq:gpbc_k}
\end{equation}
where we define the reciprocal space vectors $\vec{k} = 2 \pi \vec{n} / L$, and we recognize that 
\begin{equation}\label{eq:intft}
    \Hat{G}_{ij}(\vec{k}=\vec{0}, \omega) = \int {\rm d}t \int_{\mathbb{R}^3} {\rm d}\vec{r} \, {\rm e}^{i \omega t} G_{ij}(\vec{r}, t) = \int_{\mathbb{R}^3} {\rm d}\vec{r} \, \Tilde{G_{ij}}(\vec{r}, \omega)\,.
\end{equation}
In the main text of this Letter, we simplified the calculation by directly considering the velocity field at the origin $\vec{r}=\vec{0}$. However, the Green's function diverges at $\vec{r}=\vec{0}$ in real space. This problem is avoided in Eq.~\ref{eq:gcorr} since the $\vec{n}=\vec{0}$ term is excluded from the sum. In the following, we will keep the $\vec{r}$ dependence of the velocity field and take the limit for $\vec{r} \to \vec{0}$ later on.
The correction in Eq.~\ref{eq:gcorr} thus reads
{\small
\begin{align}
    \Delta \Tilde{G}^{\rm corr}(\omega) &= \lim_{\vec{r}\to 0} \left\{ \frac{1}{3}{\rm Tr} \left[\sum\limits_{\vec{n}, \vec{n}\neq\vec{0}} \Tilde{G}_{ij}(\vec{r} + \vec{n}L, \omega) - \frac{1}{L^3} \int_{\mathbb{R}^3} {\rm d}\vec{r}' \, \Tilde{G}_{ij}(\vec{r}', \omega) \right] \right\} \nonumber \\
    \Delta \Tilde{G}^{\rm corr}(\omega) &= \lim_{\vec{r}\to 0} \left\{ \frac{1}{3}{\rm Tr} \left[\sum\limits_{\vec{n}, \vec{n}\neq\vec{0}} \Tilde{G}_{ij}(\vec{r} + \vec{n}L, \omega) - \frac{1}{L^3}\Hat{G}_{ij}(\vec{k}=\vec{0}, \omega) \right] \right\} \nonumber \\
    \Delta \Tilde{G}^{\rm corr}(\omega) 
    &= \lim_{\vec{r}\to 0} \left\{ \frac{1}{3}{\rm Tr} \left[\frac{1}{L^3} \sum\limits_{\vec{k}, \vec{k}\neq\vec{0}} {\rm e}^{i \vec{k}\cdot\vec{r}} \Hat{G}_{ij}(\vec{k}, \omega) - \Tilde{G}_{ij}(\vec{r}, \omega) \right] \right\} \label{eq:gcorr_k} \,,
\end{align}}
where we used Eqs.~\ref{eq:gpbc_k} and \ref{eq:intft}.
Using Appendix~\ref{app:erbas}, we thus write the transverse correction as
\begin{align}
    \Delta \Tilde{G}^{T, \rm corr}(\omega) 
    &= \frac{1}{6\pi\eta}\lim_{\vec{r}\to 0} \left[\sum\limits_{\vec{k}, \vec{k}\neq\vec{0}} \frac{4\pi{\rm e}^{i\vec{k}\cdot\vec{r}}}{L^3 (k^2 + \alpha^2)} - \frac{{\rm e}^{-\alpha \vert \vec{r} \vert}}{\vert \vec{r} \vert}\right] \,. \label{eq:gcorr_k_fin}
\end{align}
We rewrite the first term in Eq.~\ref{eq:gcorr_k_fin} as
\begin{equation}
    \Tilde{f}(\vec{r},\omega) = \sum\limits_{\vec{k}, \vec{k}\neq\vec{0}} \frac{4\pi{\rm e}^{i\vec{k}\cdot\vec{r}}}{L^3 (k^2 + \alpha^2)}
    = \frac{4\pi}{L^3}\sum\limits_{\vec{k}, \vec{k}\neq\vec{0}} {\rm e}^{i\vec{k}\cdot\vec{r}} \int_0^{\infty}{\rm e}^{-(k^2+\alpha^2)x}{\rm d}x \,.
\end{equation}
The integral is then separated in two integrals from 0 to $c$ and from $c$ to $\infty$, with $c>0$ an arbitrary constant. The long range part is analytically integrated as
\begin{align}
    \Tilde{f}^{LR}(\vec{r},\omega) &= \frac{4\pi}{L^3}\sum\limits_{\vec{k}, \vec{k}\neq\vec{0}} {\rm e}^{i\vec{k}\cdot\vec{r}} \int_c^{\infty}{\rm e}^{-(k^2+\alpha^2)x}{\rm d}x \nonumber \\
    &= \frac{4\pi}{L^3}\sum\limits_{\vec{k}, \vec{k}\neq\vec{0}} {\rm e}^{i\vec{k}\cdot\vec{r}} \frac{{\rm e}^{-(k^2+\alpha^2)c}}{k^2+\alpha^2} \\
    \lim \limits_{\vec{r} \to 0} [\Tilde{f}^{LR}(\vec{r},\omega)] &= \frac{4\pi}{L^3}\sum\limits_{\vec{k}, \vec{k}\neq\vec{0}} \frac{{\rm e}^{-(k^2+\alpha^2)c}}{k^2+\alpha^2} \,,
\end{align}
where in the last line we took the limit of $\vec{r} \to 0$.
The short range part reads
\begin{equation}
    \Tilde{f}^{SR}(\vec{r},\omega) = \frac{4\pi}{L^3}\sum\limits_{\vec{k}} {\rm e}^{i\vec{k}\cdot\vec{r}} \int_0^c{\rm e}^{-(k^2+\alpha^2)x}{\rm d}x - \frac{4\pi}{L^3 \alpha^2}(1-{\rm e}^{-\alpha^2 c})\,.
\end{equation}
The first term reads
\begin{align}
    \Tilde{f}^{SR1}(\vec{r},\omega) &= 4\pi \int_0^c {\rm e}^{-\alpha^2x} \frac{1}{L^3}\sum\limits_{\vec{k}} {\rm e}^{i\vec{k}\cdot\vec{r}} {\rm e}^{-k^2x}{\rm d}x \nonumber\\
    &= 4\pi \int_0^c {\rm e}^{-\alpha^2x} \sum\limits_{\vec{n}} \frac{{\rm e}^{-\vert \vec{r} + \vec{n}L\vert^2 / 4x}}{(4\pi x)^{3/2}}{\rm d}x \nonumber\\
    &= \sum\limits_{\vec{n}} 4\pi \int_0^c {\rm e}^{-\alpha^2x}  \frac{{\rm e}^{-\vert \vec{r} + \vec{n}L\vert^2 / 4x}}{(4\pi x)^{3/2}}{\rm d}x\,,
\end{align}
where we used the Poisson summation formula. This integral can be solved, with $R=\vert\vec{r}+\vec{n}L\vert$, as
\begin{equation}
    \Tilde{f}^{SR1}(\vec{r},\omega) = \sum\limits_{\vec{n}} \frac{{\rm e}^{-\alpha R}}{2 R} \left[{\rm erfc}\left(\frac{R-2\alpha c}{2\sqrt{c}} \right) + {\rm e}^{2\alpha R} {\rm erfc}\left(\frac{R+2\alpha c}{2\sqrt{c}} \right) \right] \,.
\end{equation}
To take the limit $\vec{r} \to 0$, we separate the $\vec{n}=0$ case from the rest of the sum, yielding 
\begin{dmath}
    \lim_{\vec{r}\to 0} \Tilde{f}^{SR1}_{\vec{n}\neq0}(\vec{r},\omega) \\{= \sum\limits_{\vec{n}, \vec{n}=0} \frac{{\rm e}^{-\alpha nL}}{2 nL} \left[{\rm erfc}\left(\frac{nL-2\alpha c}{2\sqrt{c}} \right) + {\rm e}^{2\alpha nL} {\rm erfc}\left(\frac{nL+2\alpha c}{2\sqrt{c}} \right) \right]} \,.
\end{dmath}
The last term $\vec{n}=0$ is combined with the term ${\rm e}^{-\alpha r}/r$ in Eq.~\ref{eq:gcorr_k_fin} and gives by Taylor expansion
\begin{align}
    \Tilde{f}^{SR1}_{\vec{n}=0}(\vec{r},\omega) - \frac{{\rm e}^{-\alpha r}}{r}
    &\approx -\frac{{\rm e}^{-\alpha^2 c}}{\sqrt{\pi c}} + \alpha{\rm erfc}(\alpha \sqrt{c}) \,.&
\end{align}
We thus obtain, using the parameter $\epsilon = 1/2\sqrt{c}$,
\begin{align}\label{eq:ewaldT}
    6\pi\eta\Delta \Tilde{G}^{T, \rm corr}(\omega) 
    &= \frac{4\pi}{L^3}\sum\limits_{\vec{k}, \vec{k}\neq\vec{0}} \frac{{\rm e}^{-(k^2+\alpha^2)/(4\epsilon^2)}}{k^2+\alpha^2} \nonumber
    \\&+ \sum\limits_{\vec{n}, \vec{n}\neq0} \frac{{\rm e}^{-\alpha nL}}{2 nL} \left[{\rm erfc}\left(nL\epsilon-\frac{\alpha}{2\epsilon} \right) \right. \nonumber
    \\&\qquad \qquad + \left. {\rm e}^{2\alpha nL} {\rm erfc}\left(nL\epsilon+\frac{\alpha}{2\epsilon} \right) \right] \nonumber
    \\&-\frac{2\epsilon{\rm e}^{-\alpha^2 /(4\epsilon^2)}}{\sqrt{\pi}} + \alpha{\rm erfc}\left(\frac{\alpha}{2\epsilon}\right) \nonumber 
    \\&- \frac{4\pi}{L^3 \alpha^2}(1-{\rm e}^{-\alpha^2 /(4\epsilon^2)}) \,.
\end{align}
In the limit $\omega \to 0$ (\textit{i.e.} $\alpha \to 0$), Eq.~\ref{eq:ewaldT} gives 
\begin{align}\label{eq:ewaldTlim}
    6\pi\eta\Delta \Tilde{G}^{T, \rm corr}(\omega) &= \frac{4\pi}{L^3}\sum\limits_{\vec{k}, \vec{k}\neq\vec{0}} \frac{{\rm e}^{-k^2/(4\epsilon^2)}}{k^2} + \sum\limits_{\vec{n}, \vec{n}\neq0} \frac{{\rm erfc}\left(nL\epsilon\right)}{nL} \nonumber 
    \\&\qquad \qquad \qquad -\frac{2\epsilon}{\sqrt{\pi}} - \frac{\pi}{L^3 \epsilon^2} \,.
\end{align}
We verify that Eq.~\ref{eq:ewaldTlim} is identical to the Ewald expression of Yeh and Hummer~\cite{yeh_system-size_2004}. 

\begin{figure}
    \centering
    \includegraphics[width=0.5\textwidth]{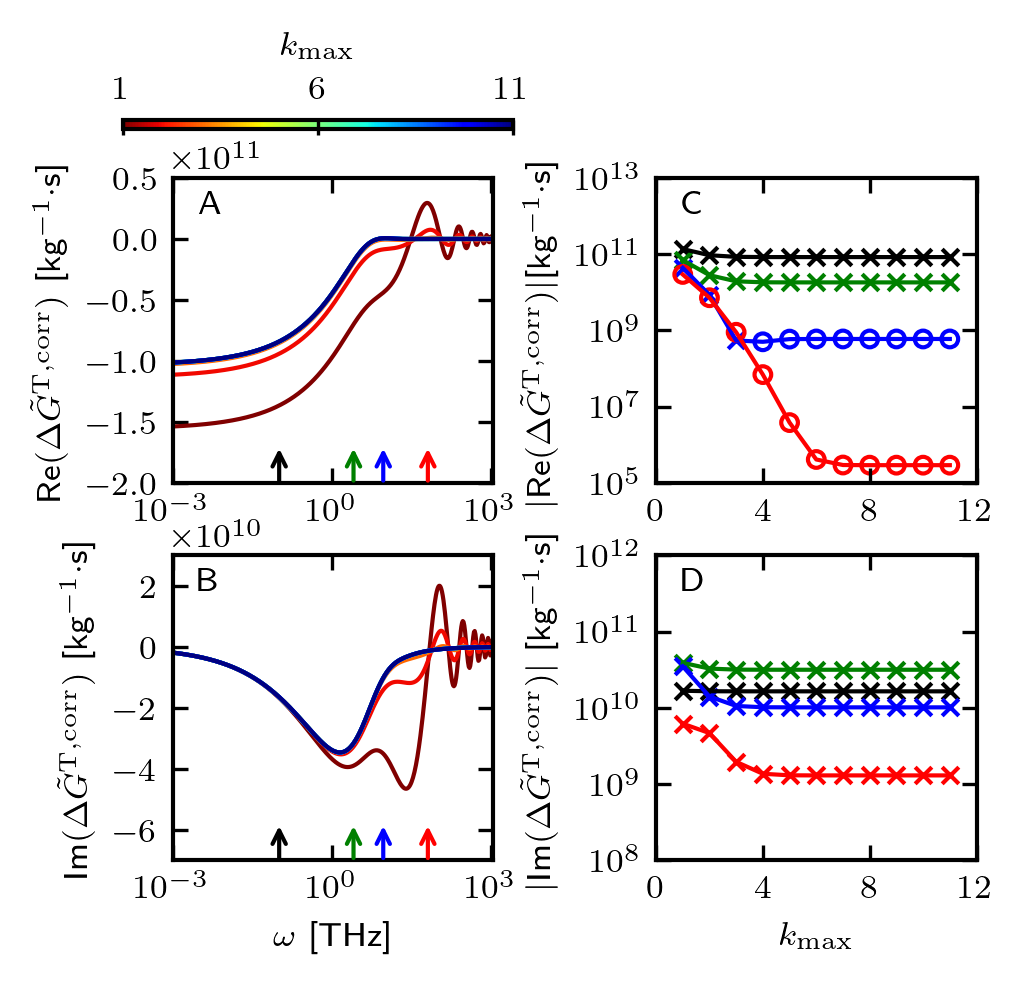}
    \caption{Test of the convergence of the Ewald summation for the transverse part Eq.~\ref{eq:ewaldT}. Frequency-dependent real (A) and imaginary (B) part of Eq.~\ref{eq:gcorr_r_fin} for different truncation indices $k_{\rm{max}}  = \llbracket 1, 2\dots11\rrbracket$, using the parameters of SPC/E water $\rho = 994$~kg.m$^{-3}$, $\eta = 0.697 \cdot 10^{-3}$~Pa.s and a box size $L=2$~nm. The sum in real space in Eq.~\ref{eq:ewaldT} is truncated at $n_{\rm max}=1$. Panels (C) and (D) show the correction in Eq.~\ref{eq:ewaldT} at $\omega=0.1,~2,~9,~65$~THz (indicated by the arrows in panels A and B) as a function of the truncation index $k_{\rm{max}}$. Crosses and circle markers indicate respectively positive and negative values.}
    \label{fig:ewald_conv}
\end{figure}

\begin{figure}
    \centering
    \includegraphics[width=0.5\textwidth]{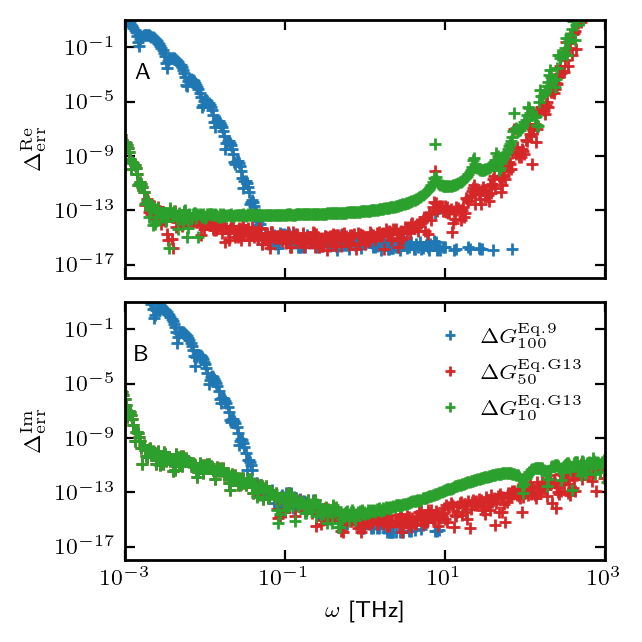}
    \caption{Relative error in the real (A) and imaginary (B) part, as defined in Eq.~\ref{eq:relerr}, for the real summation Eq.~\ref{eq:gcorr_r_fin} with $n_{\max} = 100$ (blue symbols) and for the Ewald summation Eq.~\ref{eq:ewaldT} with $k_{\rm max} = 50$ (red symbols) and $k_{\rm max} = 10$ (green symbols). For high frequencies, the error of the real space summation $\Delta_{\rm err} (\Delta G_{100}^{\rm Eq. 9})$ corresponds to the machine precision.}
    \label{fig:err_sum}
\end{figure}

Fig.~\ref{fig:ewald_conv} shows the convergence of Eq.~\ref{eq:ewaldT}, where we fixed the parameter $\epsilon = 6.5 / L$ and truncate the real space sum at $n_{\rm max} = 1$. We observe as expected that the convergence is at least 10 times faster than for the real space summation shown in Fig.~\ref{fig:real_conv} and that the small-frequency regime, where the correction is not negligible, does not show divergences. We compute relative errors with respect to the real space summation Eq.~\ref{eq:gcorr_r_fin}, which does not contain an adjustable parameter, truncated at $n_{\rm max} = 500$, given as
\begin{equation}\label{eq:relerr}
    \Delta_{\rm err}^{\rm Re}(\Delta \Tilde{G}) = \frac{\vert {\rm Re}(\Delta \Tilde{G} - \Delta \Tilde{G}^{\rm Eq. 9}_{500})\vert}{\vert {\rm Re}(\Delta \Tilde{G}^{\rm Eq. 9}_{500})\vert} \,,
\end{equation}
where we indicate the summation used in superscript and the truncation of the sum ($n_{\rm max}$ or $k_{\rm max}$ depending on the expression) in subscript. We define similarly $\Delta_{\rm err}^{\rm Im}$ for the imaginary part. Results are shown in Fig.~\ref{fig:err_sum}, where we consider only the transverse part of the corrections for the error estimates. 
We confirm that the real summation in Eq.~\ref{eq:gcorr_r_fin} is poorly converged for small frequencies for $n_{\rm max} = 100$, while the Ewald summation in Eq.~\ref{eq:ewaldT} shows good results already for $k_{\rm max} = 10$. For high frequencies ($\omega > 10$~THz), the real part of the correction goes to zero leading to a divergence of the relative error, but the absolute error remains small. 

\section{Contribution of the longitudinal part}\label{app:long}
In the case of the longitudinal contribution, using Appendix~\ref{app:erbas} and Eq.~\ref{eq:trg}, Eqs.~\ref{eq:gcorr} and \ref{eq:gcorr_k} give
\begin{align}
    \Delta \Tilde{G}^{L, \rm corr}(\omega) 
    &= \frac{\lambda^2}{12 \pi \eta \alpha^2} \left[ \sum \limits_{\vec{n}, \vec{n}\neq 0} \frac{{\rm e}^{-\lambda nL}}{nL} \right]
    - \frac{1}{3 \eta \alpha^2 L^3} \label{eq:gcorr_r_finL}\\
    &= \frac{\lambda^2}{12\pi\eta\alpha^2} \lim_{\vec{r}\to 0} \left[\sum\limits_{\vec{k}, \vec{k}\neq\vec{0}} \frac{4\pi{\rm e}^{i\vec{k}\cdot\vec{r}}}{L^3 (k^2 + \lambda^2)} - \frac{{\rm e}^{-\lambda r}}{r}\right] \label{eq:gcorr_k_finL} \,.
\end{align}
Noting the similarities between Eq.~\ref{eq:gcorr_k_finL} and Eq.~\ref{eq:gcorr_k_fin}, we derive the Ewald summation for the longitudinal contribution to the correction in a similar way as in Appendix~\ref{app:ewald}, leading to
\begin{align}
    \frac{12\pi\eta\alpha^2}{\lambda^2}\Delta \Tilde{G}^{L, \rm corr}(\omega) 
    &= \frac{4\pi}{L^3}\sum\limits_{\vec{k}, \vec{k}\neq\vec{0}} \frac{{\rm e}^{-(k^2+\lambda^2)/(4\epsilon^2)}}{k^2+\lambda^2} \nonumber
    \\&+ \sum\limits_{\vec{n}, \vec{n}\neq0} \frac{{\rm e}^{-\lambda nL}}{2 nL} \left[{\rm erfc}\left(nL\epsilon-\frac{\lambda}{2\epsilon} \right) \right.\nonumber
    \\&\qquad \qquad + \left. {\rm e}^{2\lambda nL} {\rm erfc}\left(nL\epsilon+\frac{\lambda}{2\epsilon} \right) \right] \nonumber
    \\&-\frac{2\epsilon{\rm e}^{-\lambda^2 /(4\epsilon^2)}}{\sqrt{\pi}} + \lambda{\rm erfc}\left(\frac{\lambda}{2\epsilon}\right) \nonumber
    \\&- \frac{4\pi}{L^3 \lambda^2}(1-{\rm e}^{-\lambda^2 /(4\epsilon^2)}) \,.
\end{align}
In the incompressible limit, one has $\lambda \to 0$ and the correction vanishes.

\section{Finite-size correction using frequency-dependent viscosity}\label{app:fdep}
\subsection{Calculation of frequency-dependent shear and volume viscosity spectra from MD simulations}
\label{app:green_kubo}
To investigate the influence of the frequency-dependency of the shear and volume viscosities, we calculate viscosity spectra. The shear viscosity kernel $\eta(t)$ is determined by the trace-free part of the stress tensor according to the Green-Kubo relation \cite{hansen1990theory,j2007statistical,zwanzig1965time,schulz2020molecular}
\begin{align}
\label{eq:Kubo_shear_visc}
   \Tilde{\eta}(\Vec{k} = 0, \omega) &= \int_0^{\infty}e^{i\omega t} \eta(t) {\rm d}t \nonumber \\ 
   &= \frac{V}{6 k_BT}\int_0^{\infty}e^{i\omega t} \sum_{i\neq j}\langle \Pi_{ij}(t) \Pi_{ij}(0)\rangle {\rm d}t \,,
\end{align}
where $V$ is the volume of the fluid. We define the trace-free part of the stress tensor $\sigma_{ij}$ as
\begin{equation}
    \Pi_{ij} = \sigma_{ij} - \delta_{ij}\frac{1}{3} \sum_{k} \sigma_{kk},
\end{equation}
where $i,j \in \{x,y,z\}$ .
For the computation of the shear viscosity spectrum, using Eq.~\ref{eq:Kubo_shear_visc}, we first calculate the time correlation functions of the stress tensor entries and then perform the half-sided Fourier transform.

Employing the Green-Kubo relations, we use the fluctuations of the instantaneous pressure from its average value $\langle{P}\rangle$, i.e. $\delta P(t) = P(t) - \langle P \rangle$, to compute the volume viscosity kernel $\zeta(t)$. $P(t)$ is computed from the trace of the stress tensor, i.e.  $P(t) = \frac{1}{3}\sum_{k} \sigma_{kk}(t)$. Using the half-sided Fourier transformation, we compute the volume viscosity spectrum via \cite{medina2011molecular}
\begin{align}
    \label{eq:Kubo_volume_visc}
      \Tilde{\zeta}(\Vec{k} = 0, \omega) &= \int_0^{\infty}e^{i\omega t} \zeta(t) {\rm d}t\nonumber\\ 
      &= \frac{V}{k_BT}\int_0^{\infty}e^{i\omega t} \langle \delta P(t) \delta P(0) \rangle {\rm d}t \,.
\end{align}

\subsection{Fitting of the viscosity spectra of SPC/E water}\label{app:fitting_procedure}
We apply the methods of Appendix~\ref{app:green_kubo} to MD simulations of SPC/E water~\cite{berendsen_missing_1987} in a box containing 1250 water molecules. For this, we run simulations using the GROMACS simulation package~\cite{pronk2013gromacs,abraham_gromacs_2015} (version 2020-Modified) with a time step of 2~fs. We equilibrate the system at 300~K using a Berendsen barostat~\cite{berendsen1984} at 1~atm leading to a cubic box of length 3.5616~nm. We then perform a 1~$\mu$s production run in the NVT ensemble with a temperature $T=$300~K, using a velocity rescaling thermostat~\cite{bussi2007canonical}. For electrostatics, we use the particle-mesh Ewald method~\cite{Darden_1993} with a cut-off length of 1~nm.

For practical purposes, we fit the shear and volume viscosity spectra $\Tilde{\eta}(\omega)$ and $\Tilde{\zeta}(\omega)$ extracted from the MD simulations by a combination of $N=6$ and $N=7$ exponential-oscillating functions, respectively, according to \cite{schulz2020molecular}
\begin{eqnarray}
\label{eq:visc_model_time} \nonumber
    \eta(t) = &&\Theta(t)\Bigl\{\sum_{j=I}^{N} \frac{\eta_{0,j}\tau_{n,j}}{\tau_{o,j}^2}e^{-t/2\tau_{n,j}}\Bigl[\frac{1}{\kappa_j}\sin{\Bigl(\frac{\kappa_j}{2\tau_{n,j}}t\Bigr)} \\ && + \cos{\Bigl(\frac{\kappa_j}{2\tau_{n,j}}t\Bigr)}\Bigr]\Bigr\} \,,
\end{eqnarray}
where $\kappa_j = \sqrt{4(\tau_{n,j}/\tau_{o,j})^2-1}$, which in the frequency domain becomes
\begin{equation}
\label{eq:visc_model}
    \Tilde{\eta}(\omega) =  \sum_{j=I}^{N} \eta_{0,j}\frac{1-i\omega\tau_{n,j}}{1-i\omega\tau_{o,j}^2/\tau_{n,j}-\omega^2\tau_{o,j}^2} \,.
\end{equation}
The fitting parameters are summarized in Table~\ref{tab:fit_visc}. 

\begin{table}[!htbp]
\centering
\caption{Fitting parameters for the shear viscosity $\Tilde{\eta}(\omega)$ and volume viscosity $\Tilde{\zeta}(\omega)$ in Eq.~\ref{eq:visc_model} from MD data of the SPC/E water model. The time scales are converted to frequencies.}
 \begin{ruledtabular}
 \begin{tabular}{c  c c} 

 Parameter & $\Tilde{\eta}(\omega)$ & $\Tilde{\zeta}(\omega)$ \\ 
 \hline
  $\eta_{0,I}$/$\zeta_{0,I}$ & 0.09 mPa s& 0.23 mPa s\\
   $(2\pi\cdot\tau_{n,I})^{-1 }$ & 1.89 THz & 3.46 THz \\
   $(2\pi\cdot\tau_{o,I})^{-1 }$ & 1.39 THz & 1.59 THz \\
   \hline
    $\eta_{0,II}$/$\zeta_{0,II}$ & 0.51 mPa s& 0.87 mPa s\\
   $(2\pi\cdot\tau_{n,II})^{-1 }$ & 1.73 THz & 1.32 THz \\
   $(2\pi\cdot\tau_{o,II})^{-1 }$ & 0.64 THz & 0.54 THz \\
   \hline
   $\eta_{0,III}$/$\zeta_{0,III}$ & 0.08 mPa s& 0.03 mPa s\\
   $(2\pi\cdot\tau_{n,III})^{-1 }$ & 5.64 THz & 22.77 THz \\
   $(2\pi\cdot\tau_{o,III})^{-1 }$ & 8.12 THz & 17.55 THz \\
   \hline
     $\eta_{0,IV}$/$\zeta_{0,IV}$ & 0.008 mPa s& 0.05 mPa s\\
   $(2\pi\cdot\tau_{n,IV})^{-1 }$ & 8.98 THz & 4.93 THz \\
   $(2\pi\cdot\tau_{o,IV})^{-1 }$ & 14.84 THz & 5.37 THz \\
   \hline
     $\eta_{0,V}$/$\zeta_{0,V}$ & 0.005 mPa s& 0.005 mPa s\\
   $(2\pi\cdot\tau_{n,V})^{-1 }$ & 15.87 THz & 3.77 THz \\
   $(2\pi\cdot\tau_{o,V})^{-1 }$ & 21.86 THz & 7.56 THz \\
   \hline
     $\eta_{0,VI}$/$\zeta_{0,VI}$ & 0.0008 mPa s& 0.32 mPa s\\
   $(2\pi\cdot\tau_{n,VI})^{-1 }$ & 17.79 THz & 4.22 THz \\
   $(2\pi\cdot\tau_{o,VI})^{-1 }$ & 39.03 THz & 3.16 THz \\
    \hline
     $\eta_{0,VII}$/$\zeta_{0,VII}$ & - & 0.24 mPa s\\
   $(2\pi\cdot\tau_{n,VI})^{-1 }$ & - & 0.35 THz \\
   $(2\pi\cdot\tau_{o,VI})^{-1 }$ & - & 0.14 THz \\
  \end{tabular}
   \end{ruledtabular}
 \label{tab:fit_visc}
\end{table}

\subsection{Comparison of finite-size correction using constant and frequency-dependent viscosity}

\begin{figure*}[hbt!]
    \centering
    \includegraphics[width=1\textwidth]{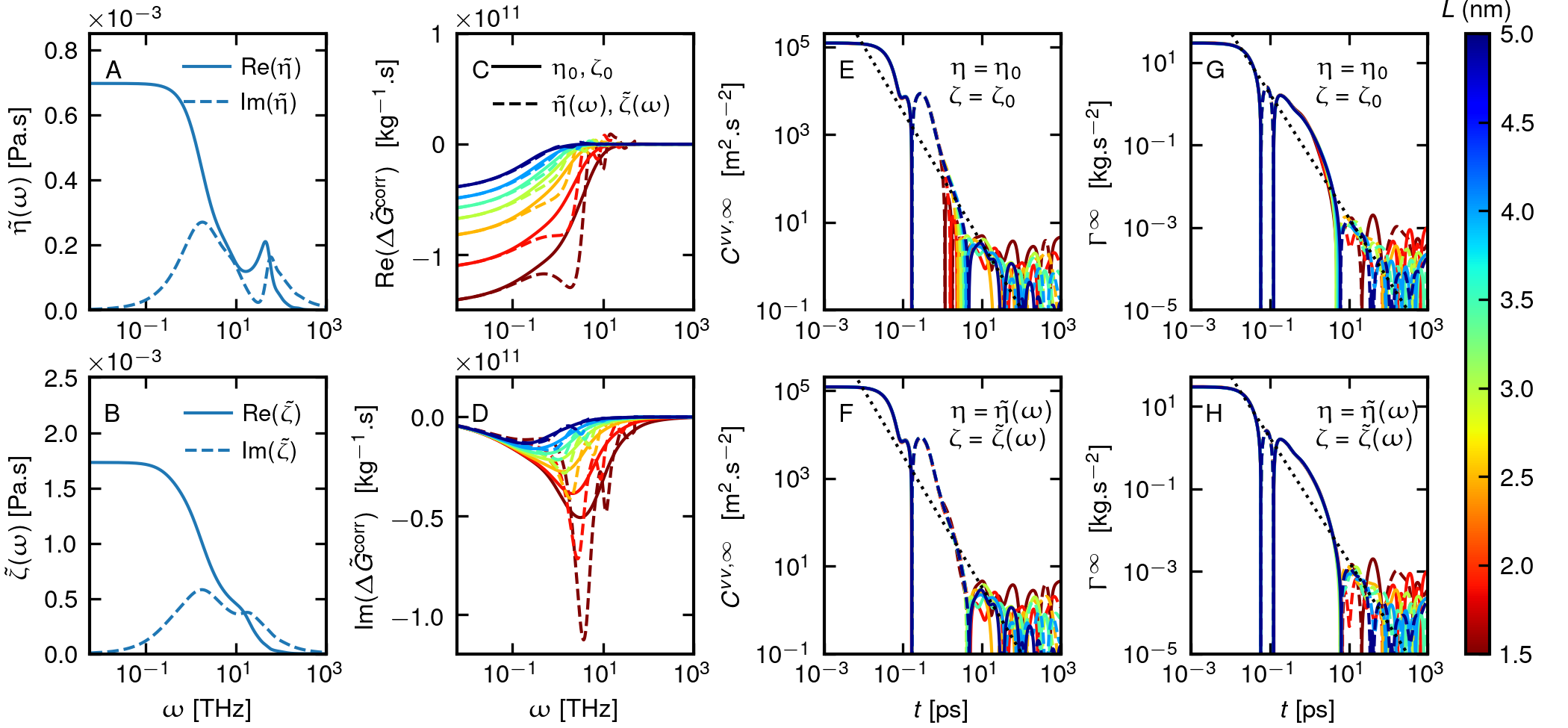}
    \caption{Fits of the frequency-dependent shear viscosity $\Tilde{\eta}$ (A) and volume viscosity $\Tilde{\zeta}$ (B). The real part is given as a solid line, while the imaginary part is dashed. Real (C) and imaginary (D) parts of the frequency-dependent finite-size correction $\Delta \Tilde{G}^{\rm corr}$ as a function of frequency, for the viscosity and density of SPC/E water, for a range of box sizes $L \in [1.5, 5]$~nm shown in the colorbar. We show the total correction including transverse and longitudinal contributions $\Delta \Tilde{G}^{\rm corr}$ in Eq.~\ref{eq:correction} computed using a constant viscosity $\eta = \eta_0$ and $\zeta = \zeta_0$ (solid lines) and computed using the frequency-dependent viscosity $\eta = \Tilde{\eta}(\omega)$ and $\zeta = \Tilde{\zeta}(\omega)$ (dashed lines). Velocity autocorrelation functions $C^{vv,\infty}$ (E-F) corrected with Eq.~\ref{eq:cvv_corr} and memory kernels (G-H) corrected with Eq.~\ref{eq:correction}, using a constant viscosity (E, G) or using a frequency-dependent viscosity (F, H). Data is shown for the center of mass position of an SPC/E water molecule in water. Dashed lines in log-log plots indicate negative values and the data is smoothed using a Gaussian filter in log space (the same filter has been applied to all data sets). The dotted lines are power-law decays $t^{-3/2}$ as predicted by the long-time tail in  Eq.~\ref{eq:gamma_tail}.}
    \label{fig:fdep}
\end{figure*}

The derivation in the main text holds also if one considers explicitly the frequency-dependence of the shear and volume viscosities. In Fig.~\ref{fig:fdep}A-B, we show the frequency-dependent fits of the shear viscosity $\tilde{\eta}(\omega)$ and volume viscosity $\tilde{\zeta}(\omega)$ computed in Appendix~\ref{app:green_kubo} and \ref{app:fitting_procedure}. The comparison of the finite-size correction Eq.~\ref{eq:correction} calculated using constant (solid lines) and frequency-dependent (dashed lines) viscosities is shown in Fig.~\ref{fig:fdep}C-D. New oscillating features appear in the terahertz regime. Note that the differences introduced by the use of the frequency-dependent viscosities are much larger than the longitudinal contribution.

We compare in Fig.~\ref{fig:fdep}E-H the VACF and the memory kernels corrected using constant (panels E and G) and frequency-dependent (panels F and H) viscosities. We observe a significant improvement of the correction using the frequency-dependent viscosity, demonstrated by the fact that the superposition of the curves from different box sizes is better. 
Note that the fitted values for the viscosity at large frequencies are very small and lead to significant numerical errors and divergences. Since the correction $\Delta \Tilde{G}^{\rm corr}$ goes to zero for large frequencies, we solve these numerical instabilities by setting the values of $\Delta \Tilde{G}^{\rm corr}$ to zero for frequencies above 50 THz.

\section{Hydrodynamic correction for a spherical geometry}\label{app:spherical}

\begin{figure}[hbt!]
    \centering
    \includegraphics[width=0.2\textwidth]{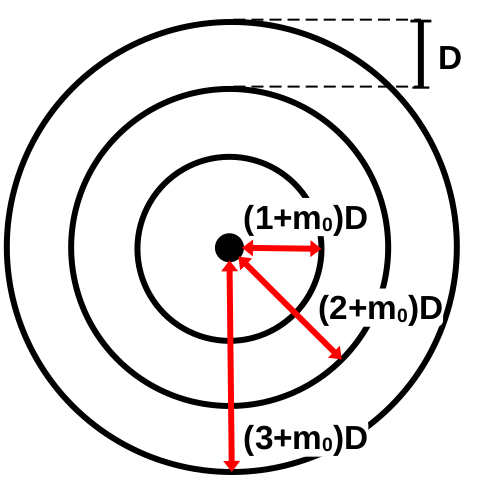}
    \caption{Spherical shell model, with adjustable parameters $D$ and $m_0$.}
    \label{fig:sph_shell}
\end{figure}
Consider the system schematized in Fig.~\ref{fig:sph_shell}: we take a set of concentric spheres so that the radius of the $m^{\rm th}$ sphere is $R_m = (m + m_0)D$, starting at $m=1$. We now take the force as $\Hat{F}_j(\vec{r}, \omega) = \left( \sum\limits_{m=1}^\infty [D \delta(\vert \vec{r} \vert - R_m)] -1 \right) \Tilde{F}_j(\omega) / V$, where we set $V = L^3$ to obtain the same force density as in the cubic case.
We then adapt Eqs.~\ref{eq:vPBCgen} and \ref{eq:gcorr} in spherical coordinates and using Eqs.~\ref{eq:trg} we write
\begin{align}
    \Delta \Tilde{G}^{T, \rm sph}(\omega) &= \frac{1}{3}
    \int_0^\infty {\rm d}r \int_0^\pi {\rm d}\theta \int_0^{2\pi} {\rm d}\phi r^2 \sin \theta \nonumber \\
    &\qquad \qquad \times \left(\sum\limits_{m=1}^\infty [D\delta(r-R_m)] - 1\right) \frac{{\rm Tr}[\Hat{G}^T_{ij}(\vec{r})]}{L^3} \nonumber \\
    &= \frac{D}{L^3}\sum\limits_{m=1}^\infty
    \frac{2 R_m {\rm e}^{-\alpha R_m}}{3\eta} - \frac{2}{3 \eta \alpha^2 L^3}\,,
\end{align}
and similarly for the longitudinal contribution
\begin{align}
    \Delta \Tilde{G}^{L, \rm sph}(\omega)
    &= \frac{D}{L^3}\sum\limits_{m=1}^\infty
    \frac{\lambda^2 R_m {\rm e}^{-\alpha R_m}}{3\eta \alpha^2} - \frac{1}{3 \eta \alpha^2 L^3}\,,
\end{align}
where in both cases we recognize the last term to be due to the background force.
Inserting the expression of $R_m$ gives for the transverse part
\begin{align}
    \Delta \Tilde{G}^{T, \rm sph}(\omega) &= \frac{2D}{3\eta L^3}\left[ \sum \limits _{m=1}^\infty (m+m_0)D {\rm e}^{-(m+m_0)\alpha D} \right] - \frac{2}{3\eta \alpha^2 L^3} \nonumber \\
    \Delta \Tilde{G}^{T, \rm sph}(\omega) & = \frac{2D^2}{3\eta L^3} \left[ \frac{{\rm e}^{-\alpha Dm_0}[-m_0+e^{\alpha D}(1+m_0)]}{({e}^{\alpha D}-1)^2} - \frac{1}{\alpha^2 D^2} \right]\,,
\end{align}
and for the longitudinal part
\begin{align}
    \Delta \Tilde{G}^{L, \rm sph}(\omega) & = \frac{D^2}{3\eta L^3} \left[ \frac{\lambda^2}{\alpha^2}\frac{{\rm e}^{-\lambda Dm_0}[-m_0+e^{\lambda D}(1+m_0)]}{({e}^{\lambda D}-1)^2} - \frac{1}{\alpha^2 D^2} \right]\,.
\end{align}
The full expression is then given by
\begin{align}
    \Delta \Tilde{G}^{\rm sph}(\omega) 
    &= \frac{D^2}{3\eta L^3} \left[ 
    \frac{2{\rm e}^{-\alpha Dm_0}[-m_0+e^{\alpha D}(1+m_0)]}{({e}^{\alpha D}-1)^2} \right. \nonumber\\
    &\left. + \frac{\lambda^2}{\alpha^2}\frac{{\rm e}^{-\lambda Dm_0}[-m_0+e^{\lambda D}(1+m_0)]}{({e}^{\lambda D}-1)^2} \right]
    - \frac{1}{\eta \alpha^2 L^3} \,.
\end{align}

\clearpage

\end{document}